\documentclass[useAMS,usenatbib]{mnras}
\usepackage[fleqn]{amsmath}
\usepackage{graphicx}	
\usepackage{multirow}
\usepackage{xcolor}

\newcommand{\der}{\mathrm{d}}
\newcommand{\R}{{R_{\rm f}}}

\newcommand{\p}{_{\rm p}} 
 
\newcommand{\vir}{_{\rm vir}}

\newcommand{\F}{^{\rm th}}

\newcommand{\beq}{\begin{equation}} 
\newcommand{\eeq}{\end{equation}}
\newcommand{\beqa}{\begin{eqnarray}}
\newcommand{\eeqa}{\end{eqnarray}} 
\newcommand{\cc}{_{\rm c}}

\newcommand{\cosm}{_{\rm cos}}
\newcommand{\critm}{_{\rm crit}}

\newcommand{\ti}{t_{\rm i}}

\newcommand{\modotb}{$M_\odot$\ } 
\newcommand{\modot}{M$_\odot$}

\newcommand{\lav}{\langle}
\newcommand{\rav}{\rangle}

\newcommand{\rhos}{\rho_{\rm s}} 
\newcommand{\rs}{r_{\rm s}} \newcommand{\Ms}{M_{\rm s}}
\newcommand{\rt}{r_{-2}} \newcommand{\Mt}{M_{-2}}
\newcommand{\Mcb}{$M-c$\/\ } \newcommand{\Mc}{$M-c$}
\newcommand{\Msrsb}{$\Ms-\rs$\ } \newcommand{\Msrs}{$\Ms-\rs$}
\newcommand{\Mtrtb}{$\Mt-\rt$\ } 
 
\newcommand{\Mab}{$M-\alpha$\ } 
\newcommand{\Msab}{$\Ms-\alpha$\ }

\begin{document}

\title[Analytic Mass-Scale Relations]{Accurate Analytic Mass-Scale Relations for Dark Matter Haloes of all Masses and Redshifts}

\author[Eduard Salvador et al.]{Eduard Salvador-Sol\'e$^1$\thanks{E-mail: e.salvador@ub.edu}, Alberto Manrique$^1$, David Canales$^2$, 
and Ignacio Botella$^3$
  \\$^1$Inst. de Ci\`encies del Cosmos i Dept. de Física Quàntica i Astrof\'{i}sica, Univ. de Barcelona. 
  Mart{\'\i} i Franqu\`es 1, E-08028 Barcelona, Spain
  \\$^2$Aerospace Engineering Department, Embry-Riddle Aeronautical University, 1 Aerospace Boulevard,
 Daytona Beach, Florida 32114, USA 
  \\$^3$Dept. of Astronomy, Graduate School of Science, Kyoto University, Kitashirakawa, Oiwakecho, Sakyo-ku, Kyoto, 606-8502, Japan}


\maketitle
\begin{abstract}
CUSP is a powerful formalism that recovers, from first principles and with no free parameter, all the macroscopic properties of dark matter haloes found in cosmological N-body simulations and unveils the origin of their characteristic features. Since it is not restricted by the limitations of simulations, it covers the whole mass and redshift ranges. In the present Paper we use CUSP to calculate the mass-scale relations holding for halo density profiles fitted to the usual NFW and Einasto functions in the most relevant cosmologies and for the most usual mass definitions. We clarify the origin of these relations and provide accurate analytic expressions holding for all masses and redshifts. The performance of those expressions is compared to that of previous models and to the mass-concentration relation spanning more than 20 orders of magnitude in mass at $z=0$ obtained in recent simulations of a 100 GeV WIMP universe.
\end{abstract}

\begin{keywords}
cosmology: theory --- dark matter --- methods: analytical 
\end{keywords}


\section{Introduction}\label{intro}

Dark Matter (DM) haloes play a central role in cosmology. Unfortunately, owing to the difficulty of treating analytically DM clustering in the highly non-linear regime, their study has so far been carried out by means of $N$-body simulations with limited mass and force resolutions. 

This is the reason that, besides a few studies on dwarf haloes with $M\sim 5\times 10^8$ \modotb at $z=3-4$ \citep{Mea01,Cea04,Iea13} and microhaloes with $M\sim 10^{-5}$ \modotb and redshifts $z=31-32$ \citep{AD13,I14}, halo density profiles have for a long time been determined for objects with masses $M\ga 10^{10}$ \modotb and redshifts $z\la 2$ (see references below). Only very recently, simulations have managed to cover haloes with masses spanning 6 orders of magnitude \citep{Iea20} and even 20 orders of magnitude \citep{Wea20b} (hereafter WBFetal) at $z=0$. 

The usual way to deal with the halo density profiles drawn from simulations is by fitting them to the NFW \citep{NFW95} or Einasto \citep{E65} parametric functions. Even though the fits are not perfect \citep{Zetal09,Mea11} and introduce spurious effects (\citealt{Sea05}), this procedure is sufficient for most purposes. One important result of that treatment is that the halo density profile appears to be universal except for the mass and redshift dependence of only one parameter \citep{NFW96}. A big effort has thus gone to determine those dependencies through the so-called mass-concentration, \Mc, or characteristic mass-scale radius, \Msrs, relations dependent on $z$. 

Simple analytic expressions, called `toy models', were put forward that fit the relations found in simulations for haloes with $M\ga 10^{10}$ \modotb and $z\la 3$ (e.g. \citealt{AR99,Cea04,Dea04,Gea08,Duf08,Mea08,Zetal09,Mea11,Kea11,PEtal12,Vea14,DM14,SP14,Hea15,Kea16,Cea18}). These toy models show, however, significant differences between authors due to the different halo samples and data treatment employed. The largest divergence is found at the high-mass end due to the different virialisation criteria used to get rid of haloes out of equilibrium (e.g. \citealt{LEtal16} and references therein). Another more technical difference between those models is that some of them  (\citealt{Bea13,Dea13,LEtal14,DK15,DJ19}; see also \citealt{PEtal12}) treat the concentration $c$ as a function of the halo seed height $\nu\equiv \delta/\sigma(M)$ instead of mass $M$. 

With the aim to go beyond the $M$ and $z$ ranges covered by simulations some `phenomenological' models were developed with a number of free parameters that were tuned through the fit to basically the same numerical data. 

The phenomenological models of first generation relied on the empirical fact that halo concentration $c$ decreases with increasing mass $M$ \citep{NFW96,Bea01,Eea01,Nea07,Mea08,DK15}. As in hierarchical cosmologies the smaller the mass of haloes, the earlier they form, that behaviour suggested that the central density of a halo should be proportional to the cosmic density at the halo formation time. Unfortunately, all these models led to an \Mcb relation of the power-law form, while later simulations showed that the real \Mcb relation flattens towards low-masses \citep{SP14,I14,LEtal16}.

The phenomenological models of second generation \citep{metal03,Sea07,LEtal14,Cea15,LEtal16,DJ19} recovered that flattening. \citet{metal03} took advantage of the fact that {\it accreting} haloes seem to grow inside-out \citep{ssm,Hea99} so that the mass accretion rate predicted e.g. in the extended-Press-Schecter (EPS) formalism \citep{PS,BCEK,B91,LC94} determined the halo density profile, with an \Mcb relation that agreed with the results of simulations \citep{Sea07}. \citet{Cea15} took the relation between the formation time and concentration found in simulations and also used the mass accretion history (MAH) of haloes predicted by the EPS formalism, while \citet{LEtal14} and \citet{LEtal16} followed the opposite scheme: they used the MAH found in simulations, which turns out to be very similar to their mass profile together with the halo formation time predicted by the EPS formalism \citep{vdB02}. (Strictly speaking, the MAH of a halo refers to its ``mass {\it aggregation} history'' rather than the mass {\it accretion} history because haloes grow not only through smooth accretion but also through major mergers.) Lastly, \citet{DJ19} relied on the observed constancy of the scale radius $\rs$ of accreting haloes at the late accreting phase \citep{Zetal03,Luea06,Sea05,Dea13b}.

Interestingly, the inside-out growth of {\it accreting haloes} assumed by \citet{metal03} is implicit in the three remaining models. Indeed, as assumed by \citet{DJ19}, \citet{Cea15} found that accreting haloes grow by keeping the scale radius unchanged as expected in inside-out growth, and the similarity between MAHs and mass profiles used by \citet{LEtal14} and \citet{LEtal16} is also implied by that growth. This suggests that the inside-out growth of accreting haloes supported by the results of simulations (e.g., \citealt{ssm,FM01,LP03,Zetal03,Sea05,Luea06,RD06,DKM07,Cea08,Wea11,LEtal13}) is crucial for the flattening of the \Mcb relation. Yet, that evolution seemed too simplistic and was actually seen as a ``pseudo-evolution'' \citep{Dea13,Wea20a}: haloes would apparently stretch outwards with increasing cosmic time even if they do not accrete simply because of the increase of the virial radius due to the decrease of the cosmic mean density. However, that argument is in contradiction with the fact that the density profile of haloes never falls off before the virial radius. Moreover, using the CUSP ({\it ConflUent System of Peak trajectories}) formalism \citep{MSS95,MSS96,Mea98}, \citet{Sea12a} (see also \citealt{SM19}) showed that the inside-out growth of accreting haloes is a natural consequence of the way accreted matter virialises.

As mentioned, the \citet{Cea15} model does not distinguish between smooth accretion and major mergers, but the other models do. In the \citet{LEtal16} and \citet{DJ19} models the effects of major mergers were taken into account in specific non-trivial manners (see Sec.~\ref{physical}). Whereas \citet{metal03} simply ignored them based on the assumption that violent relaxation causes haloes to loose the memory of their past history so that halo structure should not depend on their assembly process. That assumption seemed to contradict the ``assembly bias'' found in simulations \citep{Gea01,Gea02,ST04,FM09,FM10,Hea09,Cea20,Rea20,Hea20} suggesting that the halo density profile does depend on their merger history (e.g. \citealt{HT10,Wea20a}). However, using CUSP, \citet{SM19} have recently proven its validity, which explains the more compelling results of simulations showing that all halo properties (except for the subalo abundance, as also found by CUSP; see \citealt{II}) are independent of their assembly history \citep{WW09,Mea18}. 

Thus, CUSP confirms the validity of the \citet{metal03} model. But it does even better. It allows one to accurately derive {\it from first principles and with no single free parameter} all macroscopic properties of virialised haloes (in particular, their density profiles; \citealt{Sea12a,Sea12b,Jea14b}) from the ellipsoidal collapse and virialisation of their seeds, triaxial peaks (maxima) in the random Gaussian linear density field. It is thus much more powerful than any phenomenological model for the mass-scale relation. 

In this Paper, we use it to infer very practical, accurate and physically motivated, analytic expressions for the mass-scale relations valid for all masses and redshifts in the most relevant cosmologies and usual halo mass definitions. In Section \ref{CUSP}, we remind the derivation with CUSP of the mean spherically averaged halo density profile. Its fit to the usual NFW and Einasto analytic profiles is discussed in Section \ref{fits}. The analytic expressions for the resulting \Mcb and \Msrsb relations are given in Section \ref{analytic} and their comparison to previous models relying on the results of simulations is carried out in Section \ref{results}. The results are summarised in Section \ref{summ}.

\section{The Density Profile Predicted by CUSP}\label{CUSP}

All macroscopic properties of haloes predicted by CUSP are in very good agreement with the results of simulations. The reader is referred to \citet{SM19} for a comprehensive review of this formalism and the proofs of the two fundamental aspects of halo growth mentioned above. This is the case, in particular, of the spherically averaged density profile. Next we brievely remind how it is derived (the corresponding numerical code is available from \url{https://gitlab.com/cosmoub/cusp}).

The ellipsoidal collapse time (along all three axes) of triaxial patches at $\ti$ depends not only on their mass and size, but also on their shape and concentration (e.g. \citealt{P80}). In other words, it is a function of the density contrast $\delta$, smoothing radius $\R$, ellipticity $e$, prolateness $p$, and curvature $x$ of the corresponding peaks. However, the probability distributions functions of $e$, $p$ and $x$ of peaks with $\delta$ at $\R$ are very sharply peaked \citep{BBKS}, so all patches traced by peaks with given $\delta$ and $\R$ have essentially the same values of $e$, $p$ and $x$ and collapse at the same time. In other words, the ellipsoidal collapse time of patches essentially depends on $\delta$ and $\R$ of the peaks tracing them like in spherical collapse. Consequently, for any given $\delta(t)$ relation, we can find the radius $\R$ of the Gaussian filter such that the collapsing patches at $\ti$ traced by peaks with $\delta$ at $R$ give rise to haloes with mass $M$ at $t$. Thus, those $\delta(t)$ and $\R(M,t)$ relations establish, by construction, a one-to-one correspondence between haloes with $M$ at $t$ and peaks with $\delta$ on $\R$ at $\ti$. 

As shown by \citep{Jea14a}, these two relations, which depend on cosmology and halo mass definition, are fully determined by the consistency conditions that: 1) all the DM in the Universe at any $t$ is locked inside haloes and 2) the mass $M$ of haloes is equal to the volume-integral of their density profile. Specifically, if we write the density contrast $\delta$ for ellipsoidal collapse at $t$ and the rms density fluctuation (or 0th-order spectral moment) $\sigma_0(\R)$ of peaks in the density field at $\ti$ filtered with a Gaussian window as proportional to the homologous quantities in top-hat spherical collapse (denoted by index th), 
\beq
\delta(t,\ti)=r_\delta(t)\,\delta\F(t,\ti)
\label{deltat}
\eeq
\vspace{-15pt}
\beq
\sigma_0(\R,t,\ti)=r_\sigma(M,t)\,\sigma_0\F(\R^{\!\!\rm th},\ti)\,,
\label{rm}
\eeq
where $\R^{\!\!\rm th}=[3M/(4\pi)]^{1/3}$, $\delta\F(t,\ti)=\delta\cc\F(t)D(\ti)/D(t)$, with $\delta\F\cc(t)$ equal to the critical linearly extrapolated density contrast for spherical collapse at $t$ and $D(t)$ equal to the linear growth factor, then the numerical functions $r_\delta$ and $r_\sigma$ one is led to are well-fitted, in all cases analysed, by the simple analytic expressions, 
\beq 
r_\delta(t)\approx \frac{a^d(t)}{D(t)}
\label{cc}
\eeq
and
\beqa
r_\sigma(M,t)\!\approx \!1\!+S(t)r_\delta(t)\nu\F(M,t)~~~~~~~~~~~~~~~~~~~~~~~~~~~~~~
\nonumber\\
S(t)=s_0\!+\!s_1a(t)\!+\!\log\left[\frac{a^{s_2}(t)}{1\!+\!a(t)/A}\right],~~~~~~~~~~~~~~~~~~~~~~~~
\label{rs}
\eeqa
where $\nu\F(M,t)\equiv \delta\F(t,\ti)/\sigma\F_0(\R^{\!\!\rm th},\ti)=\delta\cc\F(t)/\sigma\F_0(M,t)$ is the (constant) linearly extrapolated top-hat height of the collapsing patch.

In Table \ref{T1} we provide the values of the coefficients in those fitting functions for the cosmologies (see Table \ref{T2}) and mass definitions used in the simulations we will compare our predictions to. Those mass definitions, which arise from the use of the Spherical Overdensity (SO) halo-finding algorithm, correspond to haloes delimited by the radius $R$ encompassing an overdensity $\Delta(z)$ relative to the characteristic cosmic density $\rho\cosm(z)$: the ``virial mass'', $M_{\rm vir}$, is for $\Delta(z)$ equal to the cosmology-dependent virial overdensity $\Delta\vir(z)$ (e.g. \citealt{bn98,H00}) and $\rho_\Delta(z)$ equal to the mean cosmic density $\rho\cosm(z)$, whereas $M_{200}$ is for a fixed value of $\Delta(z)$ equal to $200$ and $\rho_\Delta(z)$ equal to the critical cosmic density $\rho\critm(z)$.

\begin{table}
\caption{Coefficients in the halo-peak correspondence.}
\begin{center}
\begin{tabular}{cccccccc}
\hline \hline 
Cosmol. & Mass & $d$ & $s_0$ & $s_1$ & $s_2$ & $A$ \\ 
\hline 
\multirow{2}{*}
{WMAP7} & $M\vir$ & 1.06 & 0.0422 & 0.0375 & 0.0318 & 25.7\\ 
   & $M_{200}$ & 1.06 & 0.0148 & 0.0630 & 0.0132 & 12.4\\ 
\multirow{2}{*}
{Planck14} & $M\vir$ & 0.928 & 0.0226 & 0.0610 & 0.0156 & 11.7 \\ 
   & $M_{200}$ &  0.928 & 0.0341 & 0.0684 & 0.0239 & 6.87 \\
\hline
\end{tabular}
\label{T1}
\end{center}
\end{table}

\begin{table}
\caption{Cosmological Parameters.}
\begin{center}
\begin{tabular}{ccccccc}
\hline \hline 
Cosmology& $\Omega_\Lambda$ & $\Omega_{\rm m}$ & $h$ &
$n_{\rm s}$ & $\sigma_8$ & $\Omega_b$\\ 
\hline 
WMAP7 & 0.73 & 0.27 & 0.70 & 0.97 & 0.81 & 0.046\\ 
Millennium & 0.75 & 0.25 & 0.73 & 1.0 & 0.90 & 0.045\\
Planck14 & 0.68 & 0.32 & 0.67 & 0.96 & 0.83 & 0.049\\ 
\hline
\end{tabular}
\label{T2}
\end{center}
\end{table}

Differentiating with respect to $\R$ the density field smoothed with a Gaussian filter, we obtain the differential equation
\beq
\frac{\der \delta}{\der \R}=-\lav x\rav(\delta,\R)\,\sigma_2(\R) \R\,,
\label{dmd}
\eeq
where $\lav x\rav(\delta,\R)$ is the mean curvature of peaks with $\delta$ at $\R$ and $\sigma_2(\R)$ is the second order spectral moment. Given the one-to-one correspondence between haloes and peaks, $\der \delta/\der \R$ is related, through $\delta(t,\ti)$ and $M(\R,t,\ti)$ given by equations (\ref{deltat}) and (\ref{rm}), to the inverse of the instantaneous mass accretion rate of an accreting  halo and the solution $\delta(\R)$ is the continuous peak trajectory tracing its mass growth $M(t)$. 

The trajectory $\delta(\R)$ solution of equation (\ref{dmd}) determines the {\it intrinsic} (i.e. unconvolved with respect to the smoothing window) mean spherically averaged density profile, $\rho\p(r)$, of the protohalo. Indeed, taking the origin of the coordinate system at the peak on scale $\R$, the density contrast $\delta$ at ${\bf r}\p=0$ is nothing but the convolution with the Gaussian window of that radius of the (i.e. unconvolved) density contrast field $\delta\p({\bf r}\p)$ in the protohalo. That is, after integrating over the polar angles, we have
\beq 
\delta(\R)=\sqrt{\frac{2}{\pi}} \frac{1}{\R^3} \int_0^{\infty}\der r\p\,
r\p^2\,\delta\p(r\p)\,{\rm exp}\left(-\frac{r\p^2}{2\R^2}\right)\!,
\label{dp1}
\eeq 
where $\delta\p(r\p)$ is the spherical average of $\delta\p({\bf r}\p)$. Consequently, given the {\it mean} peak trajectory $\delta(\R)$ of purely accreting haloes with $M$ at $t$, by solving the Fredholm integral equation of first kind (\ref{dp1}), we can find the mean density profile $\delta\p(r\p)$ of their protohaloes \citep{Sea12a}.

Once we know the mean density profile $\delta\p(r\p)$, we can calculate the mean total energy profile 
\beqa 
\!\!\!\!E\p(r\p)\!=\!4\pi\!\!\int_0^{r\p}\!\!\!\der r\, r^2 \rho\p(r)\!\left\{\!\!\frac{\left[H_{\rm i} r\!-\!v\p(r)\right]^2}{2}\!-\!\frac{GM\p(r)}{r}\!\!\right\}~~~~\label{E1}
\\
M\p(r\p)=4\pi\int_0^{r\p} \der r\, r^2\, \rho\p(r)\,,~~~~~~~~~~~~~~~~~~~~~~~~~~~~~~~~~
\label{M1}
\eeqa
where $G$ is the gravitational constant, $\rho\p(r\p)$ stands for $\rho\cc(\ti)[1+\delta\p(r\p)]$, $H_{\rm i}$ is the Hubble constant at $\ti$ and
\beq
v\p(r\p)=\frac{2G\left[M\p(r\p)-4\pi r\p^3\rho\cc(\ti)/3\right]}{3H(\ti) r\p^2}\,
\label{vp}
\eeq
is the peculiar velocity caused by the mass excess within $r\p$.

Monitoring the ellipsoidal collapse and virialisation through shell crossing (though not apocentre crossing, which is at the base of the {\it inside-out growth} of the accreting haloes), we are led to the relation 
\beq 
r= -\frac{3}{10}\,\frac{GM^2}{E\p(M)}\,,
\label{vir0}
\eeq
between the radius $r$ and mass $M$ within it in the final virialised object (see \citealt{Sea12a} for details). Equation (\ref{vir0}) resembles the virial relation for homogeneous systems with null confining pressure, but it differs from it in that $E\p(M)$ is not the energy of the halo, $E(M)$, but that of the protohalo, which is not conserved during ellipsoidal collapse and shell crossing. Lastly, differentiating the profile $M(r)$ given by equation (\ref{vir0}), we obtain the mean spherically averaged density profile $\rho(r)$ of virialised haloes with $M$ at $t$.

\begin{figure}
\includegraphics[scale=0.44,bb= 0 160 540 690]{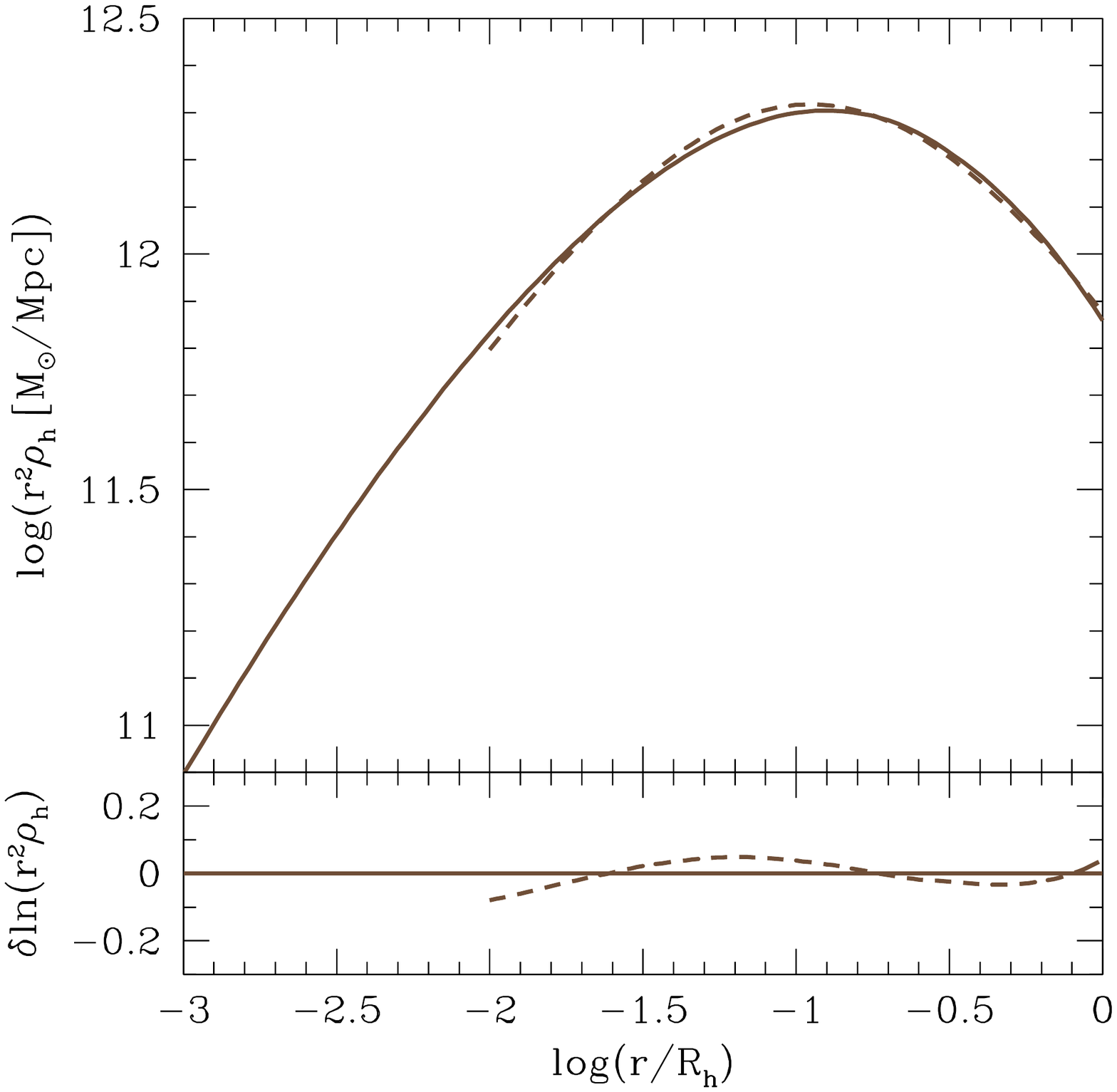}
 \caption{Mean spherically averaged density profile predicted by CUSP
   (solid line) for $z=0$ haloes with $M\vir=10^{13}$ \modotb the
   {\it WMAP7} cosmology and its unconstrained best fit to the NFW
   function down to $10^{-2} R$ (dashed line).}
\label{predict1}
\end{figure}

\begin{figure}
\includegraphics[scale=0.44,bb= 0 160 540 690]{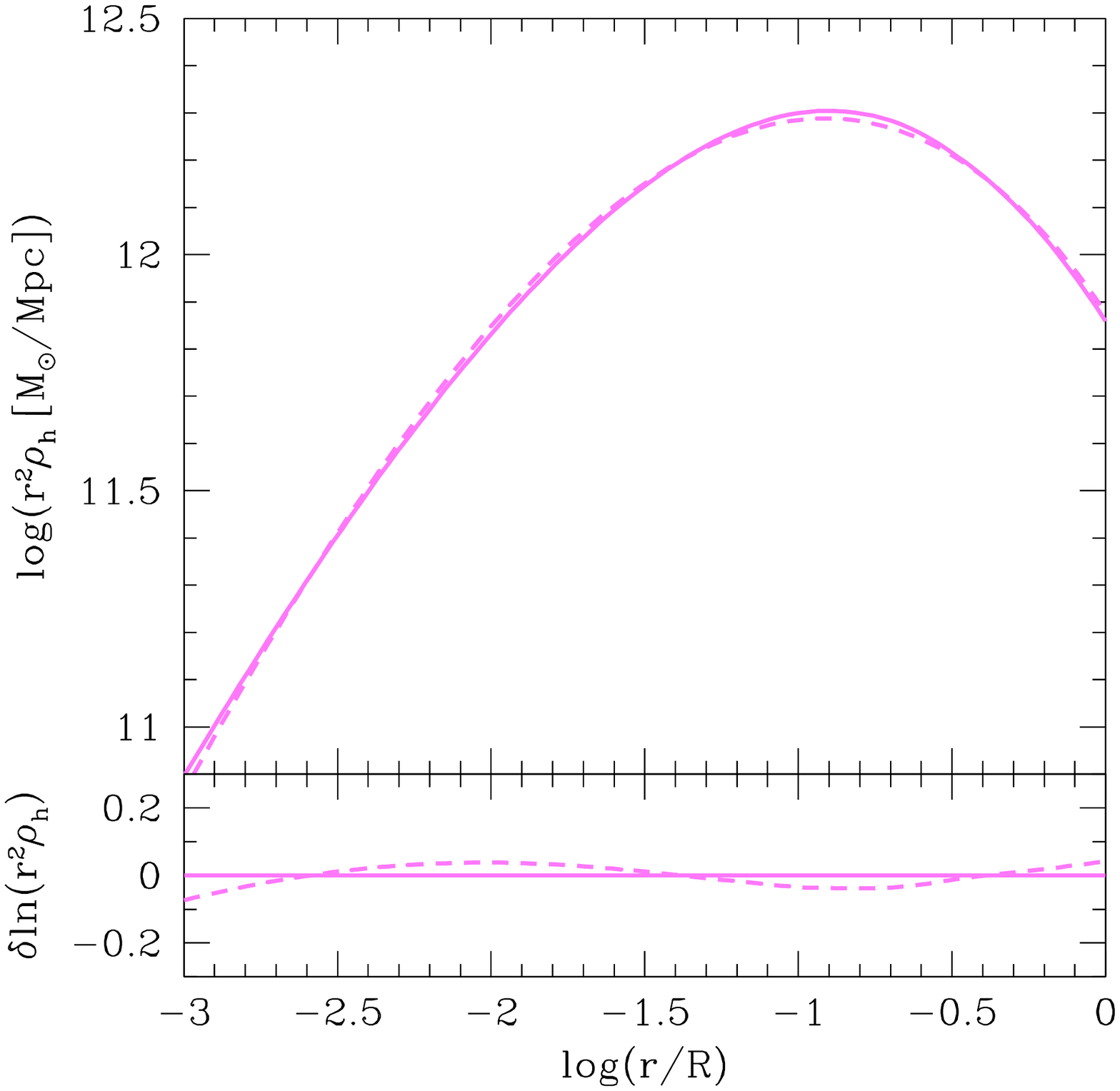}
 \caption{Same as Figure \ref{predict1} but for the Einasto fit down
   to $10^{-3} R$.}
\label{predict2}
\end{figure}

We emphasise that, even though this profile has been derived assuming purely accreting haloes, it coincides with the profile of haloes of the same mass at the same cosmic time having suffered major mergers because, as shown in \citep{SM19}, the violent relaxation suffered by haloes after a major merger causes them the memory loss of their past history. This fundamental result, formally proven in \citet{SM19}, is confirmed by the results of simulations \citep{Aea07,WW09,Mea18}. 

\section{Analytic Fits to the Density Profile}\label{fits}

This theoretical mean spherically averaged halo density profile, which is numerical, can be fitted to the usual analytic expressions used for simulated haloes, namely the two-parametric NFW profile \citep{NFW95},
\beq
  \rho(r)=\rhos\frac{4\rs^3}{r\left(r+\rs\right)^2}\,,
\label{NFW}
\eeq
and the three-parametric Einasto profile \citep{E65}, 
\beq
  \rho(r)=\rhos\exp\left\{-\frac{2}{\alpha}\left[\left(\frac{r}{\rs}\right)^\alpha-1\right]\right\}\,.
\label{Einasto}
\eeq
where $\alpha$ is the so-called shape parameter. The parameters characterising them are the scale radius $\rs$ or the concentration $c\equiv R/\rs$, where $R$ is the radius of the halo, and the characteristic density $\rhos$ or the characteristic mass within $\rs$,
\beq 
\Ms=16\pi f(1)\, \rhos\, \rs^3 ,
\label{ms1}
\eeq
with $f(x)=\ln(1+x)-{x}/(1+x)$, in the NFW case or
\beq 
\Ms=2 \pi \left(\frac{2}{\alpha}\right)^{1-{\frac{3}{\alpha}}}\,
     {\rm e}^{\frac{2}{\alpha}}\, f(1)\,\rhos\, \rs^3\,,
\label{ms2}
\eeq
with $f(x)=\Gamma(3/\alpha)-\Gamma(3/\alpha,2x^\alpha/\alpha)$, where $\Gamma(x)$ and $\Gamma(x,y)$ are the Gamma and incomplete Gamma functions, respectively, in the Einasto case. Alternatively, one can use the total mass $M$, related to $\Ms$ through
\beq
\Ms=M\,\frac{f(1)}{f(c)}\,,
\label{new}
\eeq
for the appropriate function $f(x)$ in the NFW and Einasto cases.

Note that parameters $\rs$, $\Ms$ and $\alpha$ refer to the {\it internal} structure of haloes, which is kept fixed during inside-out growth, whereas parameters $c$ and $M$ involve their {\it global} structure, which varies as haloes grow. This is the reason why the relations between the former parameters are hereafter referred to as the `internal relations' and the relations between the latter are referred to as `global relations'.

Figures \ref{predict1} and \ref{predict2} illustrate the goodness of the analytic fits to the density profiles derived by means of CUSP. As can be seen, the fits are excellent, with the residuals having the typical S-shape found in simulations (e.g. \citealt{Navea04}). Moreover, not only do the theoretical density profiles have the same shape as the empirical ones but, as we will see in Section \ref{results}, the typical values of the fitting parameters also agree.  

\begin{figure}
\includegraphics[scale=0.44,bb= 15 160 540 700]{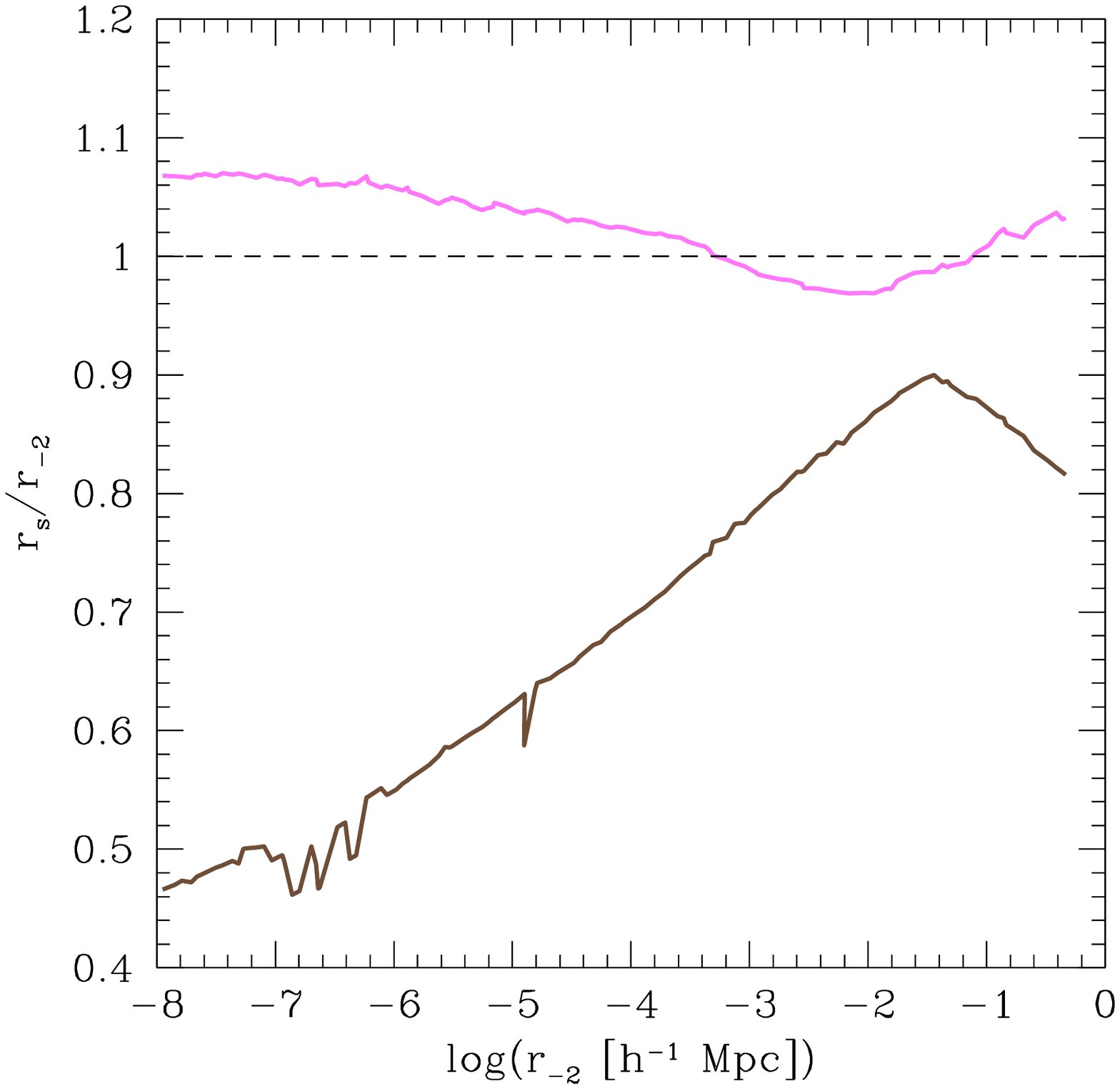}
 \caption{Best $\rs$ values found in the non-parametric fits to the NFW (lower brown line) and the Einasto (upper pink line) of the density profiles predicted by CUSP for haloes with different $r_{-2}$ values (corresponding to $M\vir$ masses spanning from $10^{-5}$ \modotb h$^{-1}$ to $10^{15}$ \modotb h$^{-1}$) in the {\it WMAP7} cosmology at $z=0$. (A colour version of this Figure is available in the online version of this Journal.)} 
\label{r-2}
\end{figure}

\begin{figure}
\vspace{-0.3cm}
\hspace{-8pt}
\includegraphics[scale=0.43,bb= 15 160 540 740]{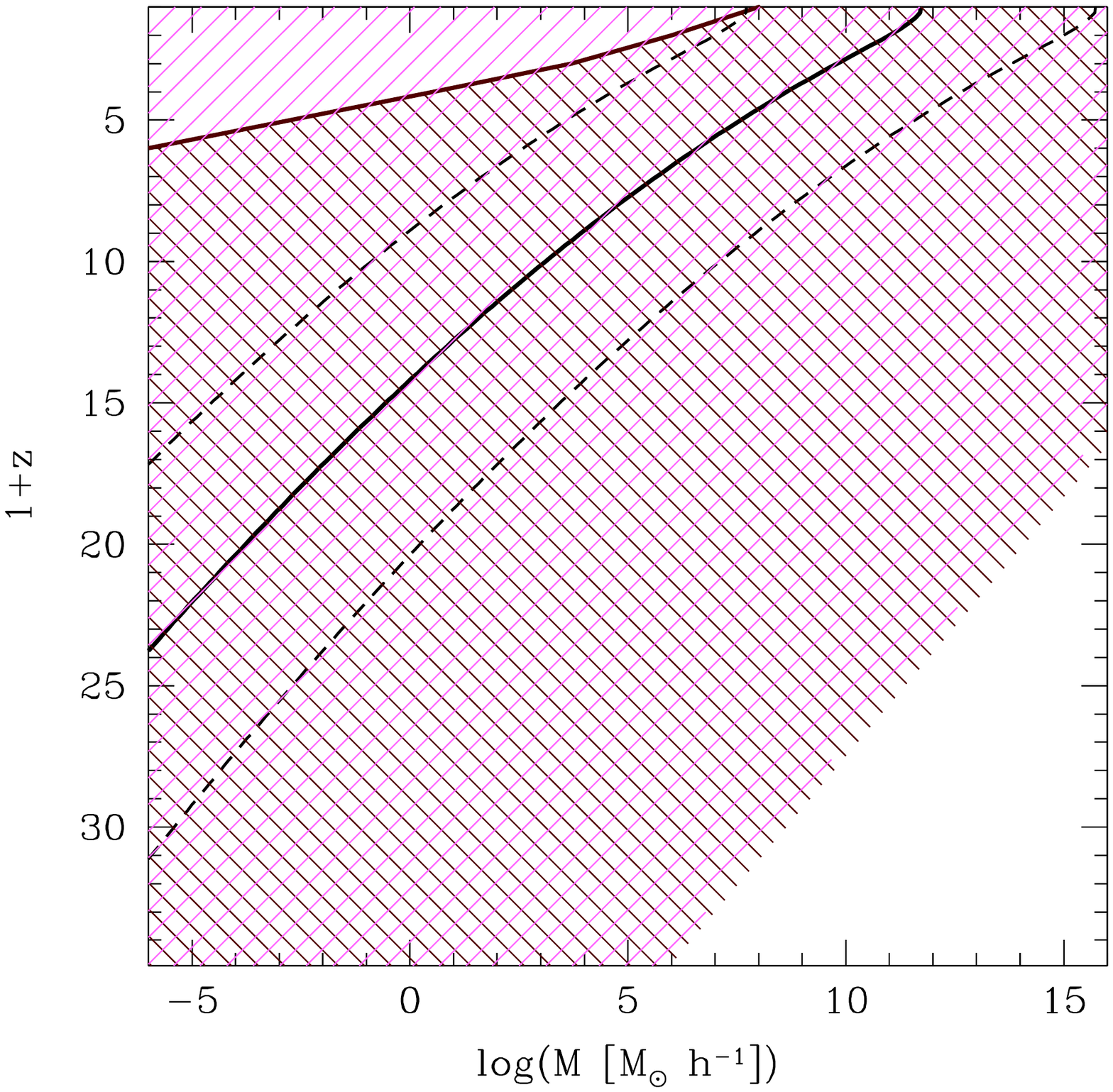}
 \caption{Domains of acceptability (according to the $\chi$-square test) of the NFW (brown upper-left to
   lower-right hatched area) and Einasto (pink upper-right to
   lower-left hatched area) fits to the density profiles
   predicted by CUSP for haloes with $M_{200}$ in the {\it
     WMAP7} cosmology with no free-streaming cutoff. The solid black line marks the $M_\ast(z)$
   curve and the dashed black lines bracket the region $10^{-4} \le M/M_\ast(z)\le 10^4$ around it. (A colour version of this Figure is available in the online version of this Journal.)}
\label{regions}
\end{figure}

In Figure \ref{r-2}, we compare the radius $r_{-2}$, where the logarithmic slope of the theoretical density profile is equal to $-2$, to the proxy $\rs$ of the best fitting NFW and Einasto functions (eqs.~[\ref{NFW}] and [\ref{Einasto}]). While in the case of the Einasto profile the difference between $r_{-2}$ and $\rs$ is small ($1.06 \ga \rs/\rt \ga 0.96$) for haloes of all masses at $z=0$ (a similar results is obtained at any other $z$), in the case of the NFW profile the solution is only acceptably good ($0.9 \ga \rs/\rt \ga 0.8$) for large halo masses ($M \ga 10^8$ \modotb h$^{-1}$). The reason for the better behaviour of the Einasto fitting function is, of course, that it involves more parameters. In the case of the NFW function $\rs$ is smaller than $\rt$ ($c$ larger than the real concentration), particularly at the low-mass end where the NFW function yields deficient fits to the very steep density profiles of low-mass haloes at low redshifts (see Fig.~\ref{regions}). In contrast, the Einasto fits are acceptable over the whole $(M,z)$ plane.  

By fitting the numerical profiles of haloes of all masses and redshifts, we have obtained the numerical dependence on $M$ and $z$ of the NFW and Einasto parameters, the so-called \Mcb and \Msrsb relations, predicted by CUSP. Since those relations will be compared to those based on the results of simulations, we have fitted the numerical density profiles inferred by CUSP as done for the density profiles found in simulations: by $\chi^2$ minimisation over the radial range from $R$ to $10^{-2} R$ with a constant logarithmic step. There are two possible ways to carry out the fits: keeping all the parameters free or enforcing their relation with the halo mass (or maximum circular velocity), which reduces the number of free parameters by one. In principle, letting all parameters free yields a better fit, but the mass of the halo with the best fitting density profile slightly differs from that of the real halo, so there is no clear advantage in any of the two procedures. In Section \ref{physical} where the CUSP-based \Mcb relation will be compared to that found by WBFetal we will carry out unconstrained fits as done by those authors. However, in Section \ref{toy} where the models our predictions will be compared to use both kinds of fits, we will adopt the geometric mean of the values obtained in the two kinds of fits. This is enough, indeed, because the relative difference between the parameter values found in the two ways is small ($< 3$\% at $10^{-4} M_\ast(z)$ and up to about 6\% at $M=10^{3}M_\ast(z)$, where $M_\ast(z)$ is the critical mass for ellipsoidal collapse at $z$ solution of the equation $\sigma[M_\ast(z),z]=\delta\cc(z)$). 

\begin{figure}
\centerline{\includegraphics[scale=0.41,bb=45 110 540 700]{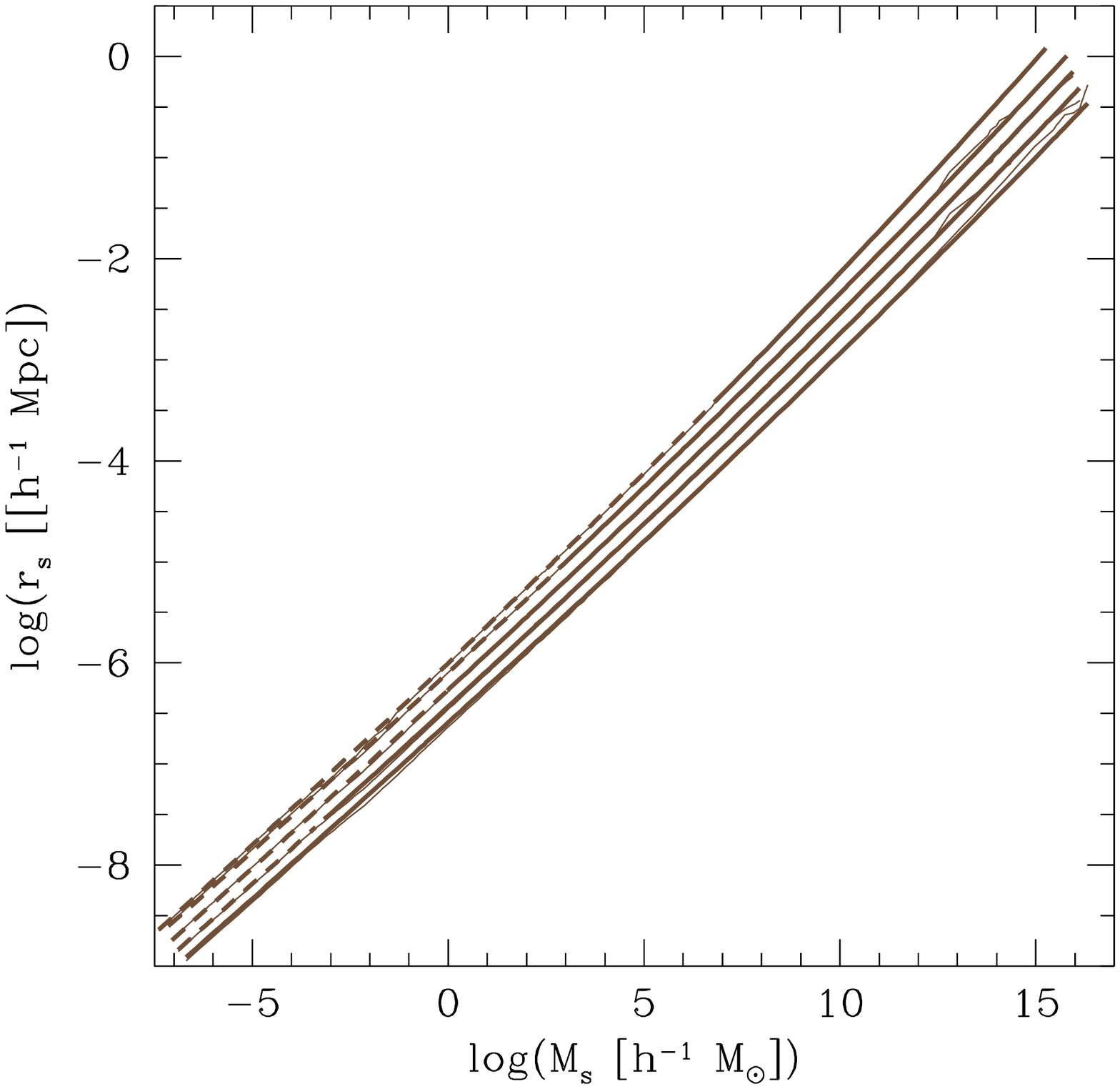}}
\centerline{\includegraphics[scale=0.41,bb=45 145 540 650]{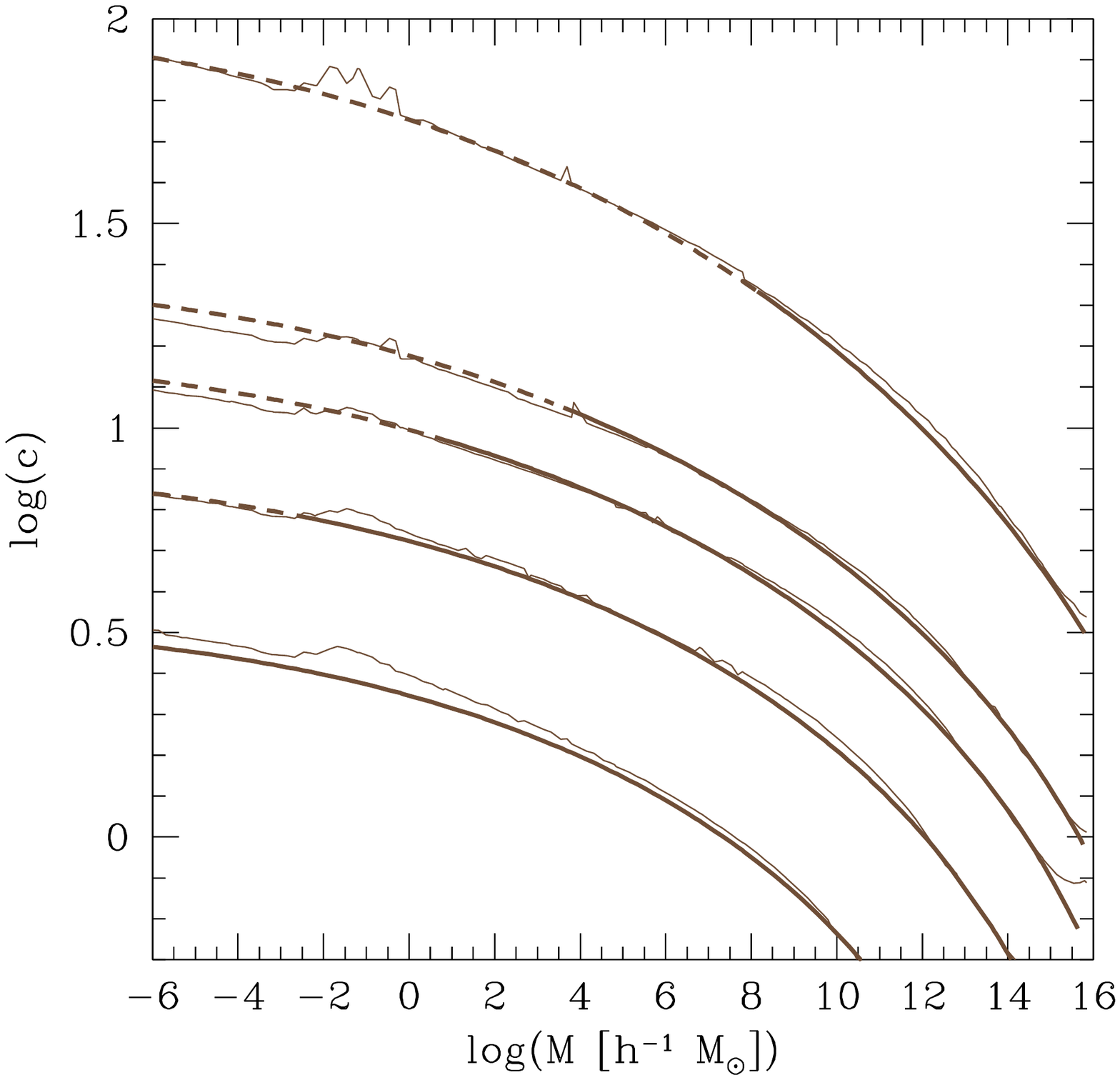}}
 \caption{{\it Top panel}: \Msrsb relations resulting from the fits to the NFW profile of the halo density profiles predicted by CUSP in the {\it WMAP7} cosmology (thin lines) and the corresponding fits to the analytic expression (\ref{power})-(\ref{nose}) (thick lines)for all relevant $M\vir$ masses and $z=0$, 2, 3 , 5 and 9 (from top to bottom). To avoid overlapping the curves for $z>0$ have been progressively shifted 0.2 dex downwards. The regions where the NFW fits are deficient are marked with dashed lines. {\it Bottom panel}: Same as the top panel, but for the \Mcb relations using the analytic expression (\ref{Mcz}). No shift has been applied to these curves.}  
\label{mc2}
\end{figure}

\begin{figure}
\centerline{\includegraphics[scale=0.41,bb= 45 110 540 700]{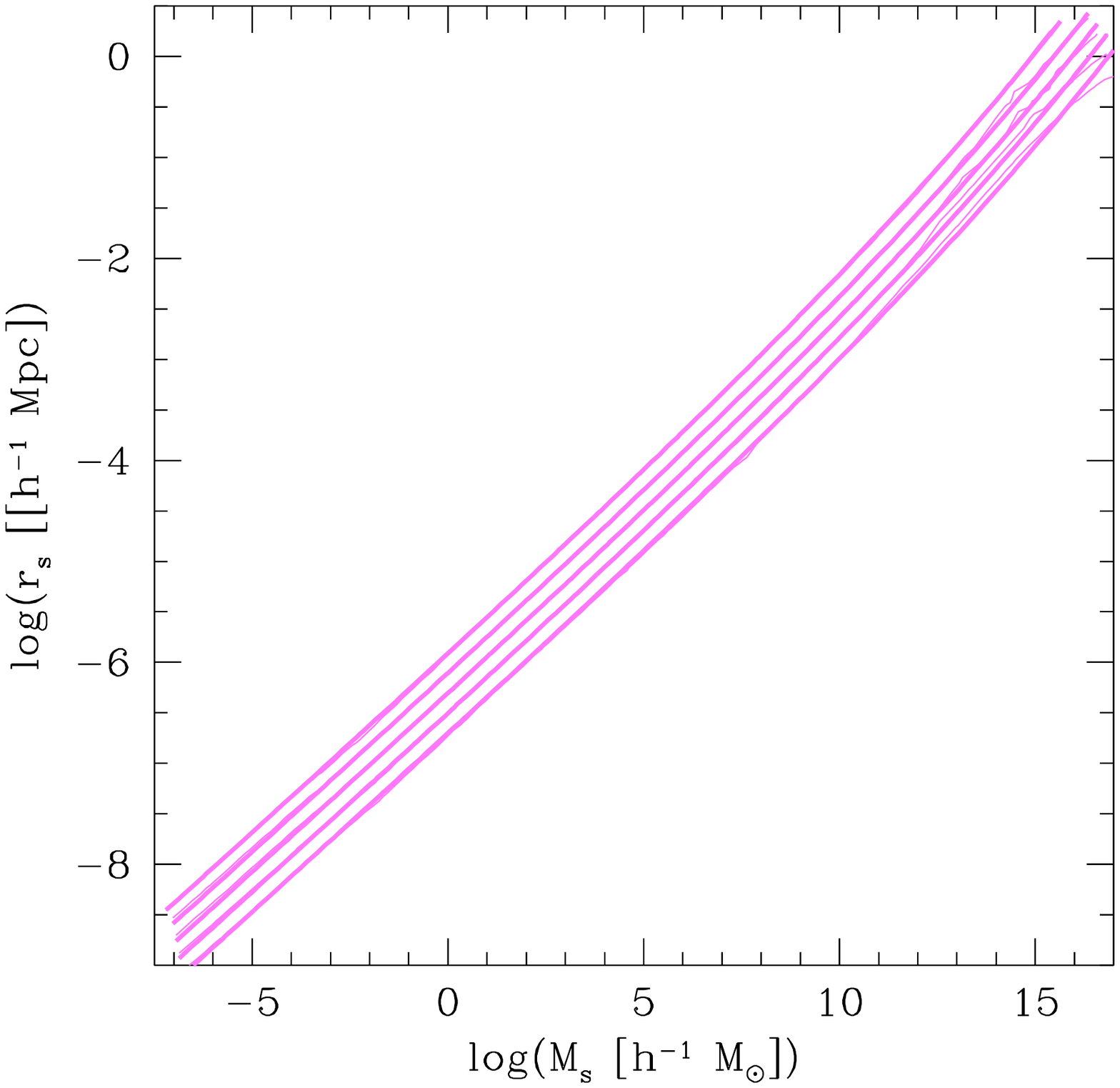}}
\centerline{\includegraphics[scale=0.41,bb= 45 150 540 650]{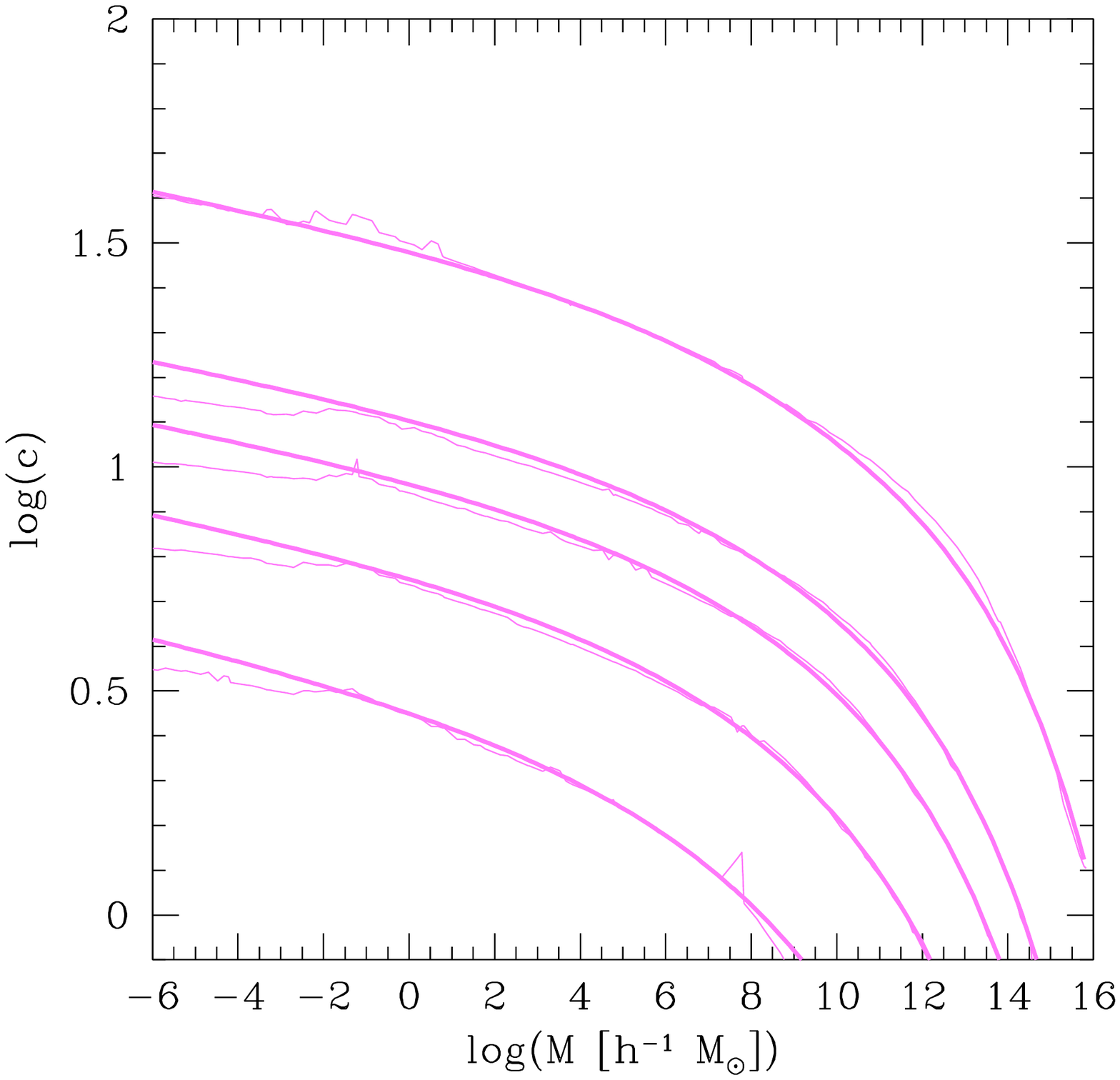}}
 \caption{Same as Figure \ref{mc2} but for Einasto profile and $M_{200}$ masses in the {\it Planck14} cosmology (thin and thick lines). {\it Top panel}: \Msrsb  relations. {\it Bottom panel}: Corresponding \Mcb relations.}
\label{E_mc2}
\end{figure}

\begin{table*}
\begin{center}
\caption{Coefficients in the NFW and Einasto \Msrsb relations.}
\begin{tabular}{cccccccccccc}
\hline \hline Cosmology & Mass & Profile & $r_0$ ($10^{-5}$ Mpc) & $M_0$ (\modot) & $\tau_0$ & $t_1$ & $t_2$ & $t_3$ & $t_4$ \\ \hline 
\multirow{4}{*}{WMAP7} &\multirow{2}{*}{$M\vir$} & NFW & 9.46 & $1.00\times 10^5$  & 0.325 & 0.183 & $-0.192$ & $-0.346$ & .0145\\ 
      &        & Einasto & 10.2 & $8.91\times 10^4 $ &  0.311 & 0.213 & $-.0234$ & $0$ & .0183\\ 
      &\multirow{2}{*}{$M_{200}$} & NFW & $9.75$ & $1.00\times 10^5$  & 0.317 & 0.199 & $-0.124$ & $-0.221$ & .0134 \\ 
      &         & Einasto & 10.8& $8.91\times 10^4 $  & 0.325 & 0.155 & $-.0325$  & 0 &  .0224\\
      \multirow{4}{*}{Planck14} &\multirow{2}{*}{$M\vir$}  & NFW & 8.04 & $1.00\times 10^5$  & 0.280 & 0.382 & $-0.113$ & $-0.349$ & .00854\\
      &          & Einasto & 8.91 & $8.91\times 10^4 $ &  0.344 & .0717 & $-0.117$ & $0$ & .0467\\  
      &\multirow{2}{*}{$M_{200}$} & NFW & 8.59 & $1.00\times 10^5$  & 0.314 & 0.219 & $-0.134$ & $-0.238$ & .0134 \\ 
      &          & Einasto & 10.0 & $8.91\times 10^4 $  & 0.353 & .0510 & $-0.100$  & 0 &  .0503\\
      \hline
      Planck14$^*$ & $M_{200}$ & Einasto & 10.1 & $8.91\times 10^4 $ & 0.347 & .0673 & --- & --- & .0388  \\
\hline
\end{tabular}
\label{T3}
\end{center}
\vspace{-5pt}
$^*$ Mass-unconstrained fit with $\alpha$ fixed according to equation (\ref{alpha2}) with coefficients given in Table \ref{T4}.
\end{table*}

\begin{table*}
\begin{center}
\caption{Coefficients in the Einasto \Msab relation.}
\begin{tabular}{cccccccccccccccc}
\hline \hline 
Cosmol. & Mass & $M_1$ (\modot) & $m$ & $\alpha_0$ & $a_{1,1}$ & $a_{1,2}$ & $a_{2,1}$ & $a_{2,2}$ &  $\alpha_3$ \\ 
\hline 
\multirow{2}{*}{WMAP7} 
& $M\vir$ & $2.63\times 10^{10}$ & $-.0648$ & .317 & $-1.275$ & .0348 & .00434 & 0.445 & $-.0523$ \\
& $M_{200}$ & $0.69\times 10^{10}$ & $-.0615$ & .290 & $-1.189$ & .0306& .00449 & 0.545 & $-.0538$ \\
\multirow{2}{*}{Planck14} 
& $M\vir$ & $7.66\times 10^{10}$ & $-.0543$ & .300 & $-1.250$ & .0457 & .00487 & 0.621 & $-.0502$ \\ 
& $M_{200}$  & $1.89\times 10^{10}$ & $-.0355$ & .300 & $-1.219$ & .0287 & .00478 & 0.455 & $-.0605$ \\ 
\hline
Planck14$^*$ & $M_{200}$ & $9.43\times 10^{9}$ & $-.0542$ & .302 & $-1.204$ & --- & $.00396$ & ---  & ---  \\
\hline
\end{tabular}
\label{T4}
\end{center}
\vspace{-5pt}
$^*$ Mass-unconstrained three-parametric fit.
\end{table*}

\section{Analytic Mass-Scale Relations}
\label{analytic}

\begin{figure}
\includegraphics[scale=0.43,bb= 15 150 540 690]{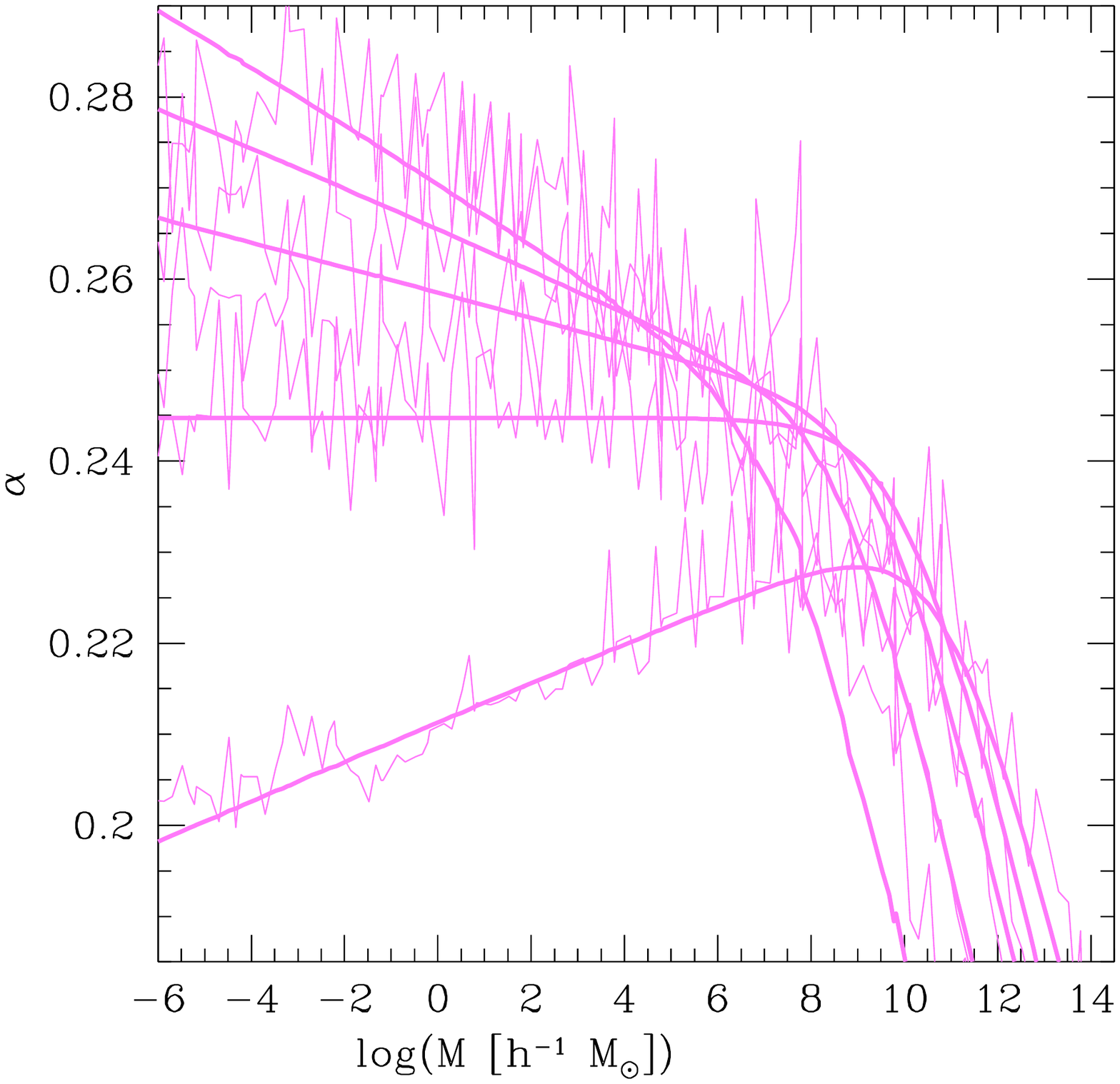}
 \caption{Same as Figure \ref{E_mc2} for the  \Mab relations. The lowest curve on the left corresponds to $z=0$, and the uppermost one to $z=9$.}
\label{E_ma2}
\end{figure}

Thus, the fact that major mergers yield halo density profiles identical to those arising from accretion \citep{SM19}, which develop inside-out, causes the typical spherically averaged density profiles of virialised haloes to be fully determined by those of peaks at $\ti$. The result is that halo density profiles are close to the NFW and Einasto form with the respective internal parameters satisfying very simple relations. 

Indeed, the \Mtrtb relation, with $M_{-2}\equiv M(r_{-2})$, is basically a power-law. The reason for this is that both $\sigma_2(\R)$ and $\lav x\rav(\delta,\R)$ are closely power-laws (in the halo mass range, the CDM power spectrum behaves as a power-law) and so is also the mean trajectory $\delta(\R)$ of accreting haloes (eq.~[\ref{dmd}]). Since the boundary condition $\delta$ at $\R(M,t)$ is also close to a power-law of $M$ (eq.~[\ref{rm}]), the whole solution $\delta(\R)$ will essentially behave as a power-law of $M$ at every fixed $\R$ too. And the same is true for the {\it unconvolved} protohalo density contrast $\delta\p(r)$ at any fixed $r$, which implies that the total energy of protohaloes $E\p(M)$ is also closely a power-law of $M$ (eqs.~[29]-[30] in \citealt{Sea12a}). Equation (\ref{vir0}) then implies that the mass $M$ inside the radius $r$ along the evolution of any accreting halo is approximately a power-law, too, with the same index for all haloes, which explains that the \Mtrtb relation in all accreting haloes approximately satisfies the same linear log-log relation. And, since the density profiles in haloes having suffered major mergers are indistinguishable from those of purely accreting haloes, the same conclusion holds for {\it all} haloes.

Therefore, since $\rs$ is a good proxy for $\rt$ (Fig.~\ref{r-2}), the \Msrsb relation must be close to a power-law,
\beq
\rs=r_0 \left(\frac{\Ms}{M_0}\right)^{\tau},
\label{power}
\eeq
with index $\tau$ independent of mass. Moreover, as $\rs$ and $\Ms$ are internal parameters, $\tau$ must also be independent of $z$. And, as can be seen by dividing equation (\ref{ms1}) by equation (\ref{ms2}), the third internal parameter, $\alpha$, must also have an approximately fixed value at least for large masses where both the NFW and Einasto profiles provide acceptable fits.

Strictly speaking, since the power-law form of $\rs$ (eq.~[\ref{power}]) is just a good approximation, $\tau$ and $\alpha$ will slightly depend on $\Ms$. Moreover, since the fit of the density profiles to the usual analytic functions is not perfect, the best fitting values of the internal parameters also slightly vary with $z$ due to the variation of the fitted radial range with halo growth \citep{Sea12a}. Consequently, we must allow for $\tau$ and $\alpha$ to slightly depend on both $\Ms$ and $z$. 

The expressions 
\beq 
\frac{\tau}{\tau_0}=1 + t_1(1+z)^{t_2}\left[\frac{\Ms}{M_0(1+z)^{t_3}}\right]^{t_4}
\label{nose}
\eeq
and
%
%
\beq
\frac{\alpha}{\alpha_0}\!=\!1 +\alpha_1(z)\left\{\! 1\! +\! \left[\frac{M\cc(z)}{\Ms}\right]^{\!\frac{1}{2}}\!\right\}^{\!\alpha_2(z)}\!\!\!+\left\{\!1\!+\!\left[\frac{\Ms}{M\cc(z)}\right]^{\!\frac{1}{2}}\!\right\}^{\alpha_3}\!,
\label{alpha2}
\eeq
with
\beqa
\alpha_1(z)= a_{1,1}\,[1-a_{1,2}\,{\rm erf}(z)]~~~~~~~~~~~~~~~~~~~~~~~~~~~~~~~~~~~~~~~~~~\nonumber\\
\alpha_2(z)=a_{2,1}(1-z/2)^{a_{2,2}}\,~~~~~~~~~~~~~~~~~~~~~~~~~~~~~~~~~~~~~~~~~~~~~\nonumber\\
\log[M\cc(z)] = \log(M_1)\exp(m\,z)\,,~~~~~~~~~~~~~~~~~~~~~~~~~~~~~~~~~~~~\nonumber
\eeqa
where erf is the error function, give excellent fits to the numerical relations predicted by CUSP. In Tables \ref{T3} and \ref{T4} we provide the best values of the coefficients for the cases of interest used in this work. 

Taking into account the definition of concentration, $c=R/\rs$, and the relation (\ref{new}), the previous internal relations lead to the \Mcb relation
\beqa 
\log (c)+\tau(M,z) \log\left[\frac{f(1)}{f(c)}\right]\!=\!\frac{1}{3} \log\left[\frac{[M/M_0]^{1-3\tau(M,z)}}{\mu(z)}\right]
\label{Mcz}\\
\mu(z)=\frac{4\pi\Delta(z)\rho_\Delta(z)\,r_0^3}{3 M_0}~~~~~~~~~~~~~~~~~~~~~~~~~~~~~~~~~~~~~~~~~~
\label{Mast}
\eeqa
with $\tau(M,z)$ given by expressions (\ref{nose}) and to the \Mab relation given by equation (\ref{alpha2}) with $\Ms$ replaced by $Mf(1)/f(c)$. Note that the \Mab relation has an extra implicit dependence on $z$ through $c$. 

The goodness of the previous analytic fitting expressions for the NFW and Einasto \Msrsb and \Mcb relations is seen in Figures \ref{mc2}, \ref{E_mc2} and \ref{E_ma2}, where they are compared to the numerical relations directly arising from the fits to the halo density profiles predicted by CUSP for different cosmologies and mass definitions (see Sec.~\ref{results}). The large oscillations found in the best fitting values of $\alpha$ (Fig.~\ref{E_ma2}) show the difficulty of determining this parameter due to the degeneracy in the three-parametric fit. Fortunately, the best values of the other two parameters, $\rs$ and $\rhos$ (or $\Ms$) are weakly dependent on the exact value of $\alpha$, so they are well determined anyway (see Fig.~\ref{E_mc2}). 

We remark that, while the \Msrsb and \Msab relations (eqs.~\ref{power}--\ref{alpha2}) are explicit for $\rs$ and $\alpha$, the \Mcb and \Mab relations are implicit for $c$ and $\alpha$, so the former are more practical than the latter. Nevertheless, in small mass ranges the dependence of $c$ on $M$ can be approximated by a simple power-law relation as found in classical toy models. Indeed, $f(c)$ is essentially constant\footnote{$f(c)$ is a smooth function of $c$, and $c$ is little dependent on $M$ because $\tau$ is close to $1/3$.} and $\tau$ is little dependent on $M$ so the \Mcb relation (\ref{Mcz}) is close to a linear log-log relation. On the other hand, the same approximations (i.e. $f(c)$ constant and $\tau\approx 1/3$) also lead to 
\beq
c \propto \left[\Delta(z)\rho_\Delta(z)\right]^{-1/3},
\label{proxi}
\eeq
implying that $c$ is roughly proportional to $(1+z)^{-1}$ in small redshift ranges as found by \citet{Bea01}, though equation (\ref{proxi}) is a better approximation. 

\section{Comparison with Previous Models}\label{results}

The comparison of those CUSP-based analytic mass-scale relations to previous toy and phenomenological models will be achieved in the two different mass and redshift regimes probed by simulations and for both the `global' and `internal' relations. These two kinds of relations are equivalent, but, while the former is the most commonly used, the latter informs more directly on halo growth. 

There are of course small differences in the data treatment and fitting procedure used by different authors (including ourselves), but they are not expected to significantly affect the comparisons. The only differences that, in principle, might substantially affect them are: 1) while the concentration obtained by means of CUSP refers to the {\it mean} density profile of haloes with a given mass, that considered in most models is the {\it median} value and 2) while CUSP deals by definition with virialised haloes, simulations include to some extent haloes out of equilibrium. Point 1 could be a problem because the concentration of haloes of a given mass is lognormally distributed with a notable scatter \citep{DM14}, implying that the mean concentration is substantially larger than the median one. However, the concentration calculated by means of CUSP is not the mean concentration of haloes with fixed mass, but the concentration of the mean density profile of those haloes and, as shown in the Appendix, this latter value coincides with the median concentration. Thus, there is no problem in this respect. Regarding point 2, we must say that the departure from equilibrium of haloes really causes the \Mcb and \Msrsb relations at the high-mass end at every redshift to differ between different models and the CUSP-based ones. Indeed, a large fraction of simulated haloes in that mass regime, where major mergers are more common, are not fully relaxed and, even though the authors enforce different virialisation criteria in order to select those which are, this objective is not fully accomplished \citep{LEtal16}. 

\subsection{High-Mass Low-redshift Regime}\label{toy}

Next we compare our analytic relations to previous toy models in the high-$M$ low-$z$ regime. 

Among all the models focusing on the NFW profile we have chosen those provided by \citet{Zetal09}, \citet{Mea11} and \citet{Kea11} since they all refer to the same {\it WMAP7} cosmology \citep{Km11} or close to it (see Table \ref{T2}) and the same $M\vir$ masses, which we also adopt for the CUSP-based relations. Other more recent toy models \citep{PEtal12,DM14,Kea16,Iea20} using other cosmologies, parametrisations or fitting techniques give similar results, however. We must also say that \citet{Zetal09} and \citet{Mea11} adjust two parameters, while \citet{Kea11} adjust only one parameter (they enforce the value of the maximum circular velocity $V_{\rm max}$). This is the reason why we use the hybrid fit technique explained in Section \ref{fits}. 

For the Einasto profile, the toy models available are those provided by \citet{Gea08} for the {\it Millennium} cosmology (with parameters close to the Planck ones; see the parameters values in Table \ref{T2}) \citep{Sprea05} and by \citet{DM14} and \citet{Kea16} for the {\it Planck14} cosmology \citep{P14,Kea16} (see Table \ref{T2}). The masses adopted in all those studies are $M_{200}$. We thus use the CUSP-based relations for that cosmology and mass definition. Again, \citet{Gea08} and \citet{DM14} used unconstrained fits, while \citet{Kea16} used constrained ones. \citet{LEtal13} repeated the study by \citet{Gea08} for the same cosmology and with more particles per halo, but they focus on $z=0$, so we compare our analytic expressions to the former model.  

\begin{figure}
\centerline{\includegraphics[scale=0.41,bb=45 150 540 700]{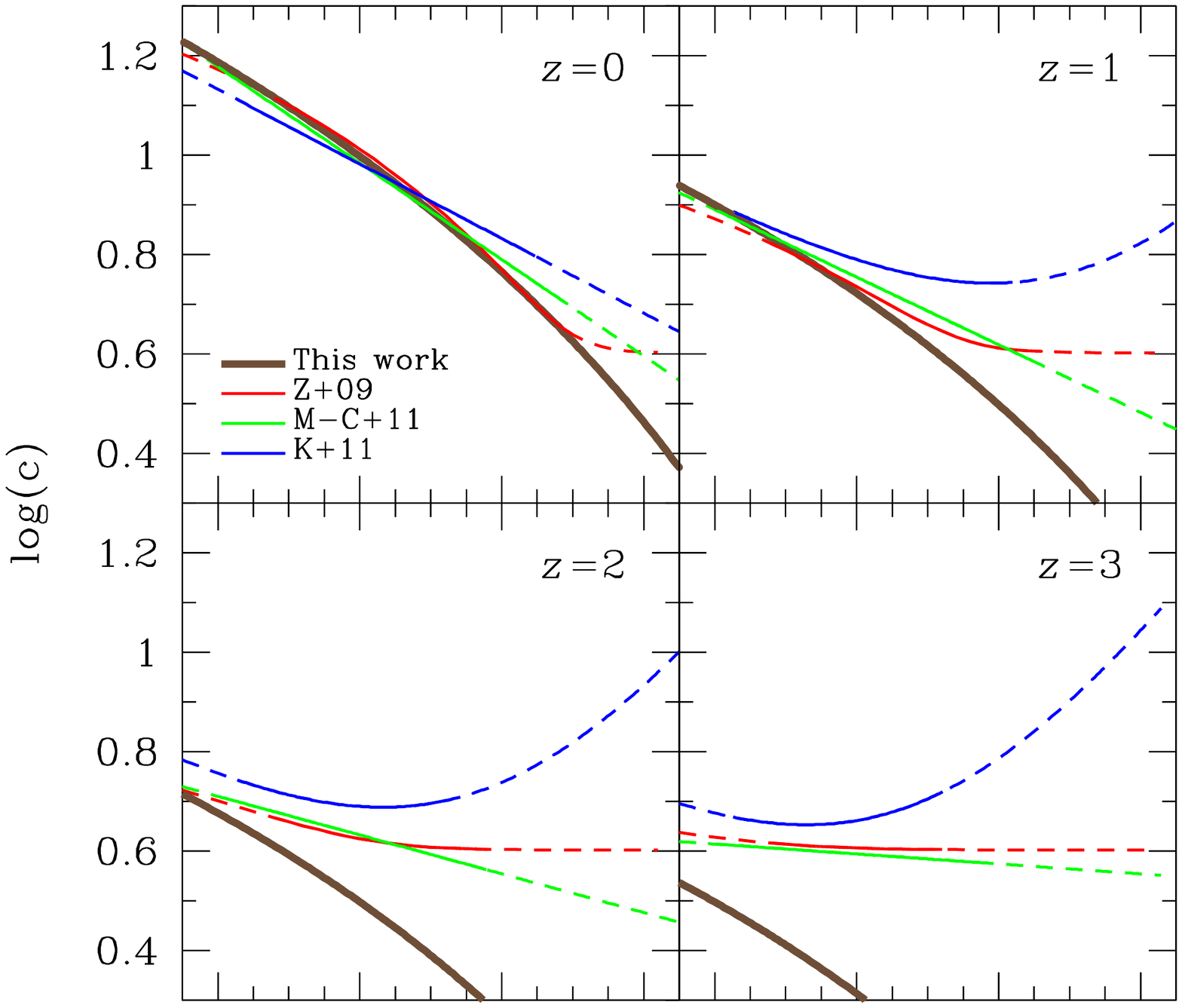}}
\centerline{\includegraphics[scale=0.41,bb=45 150 540 620]{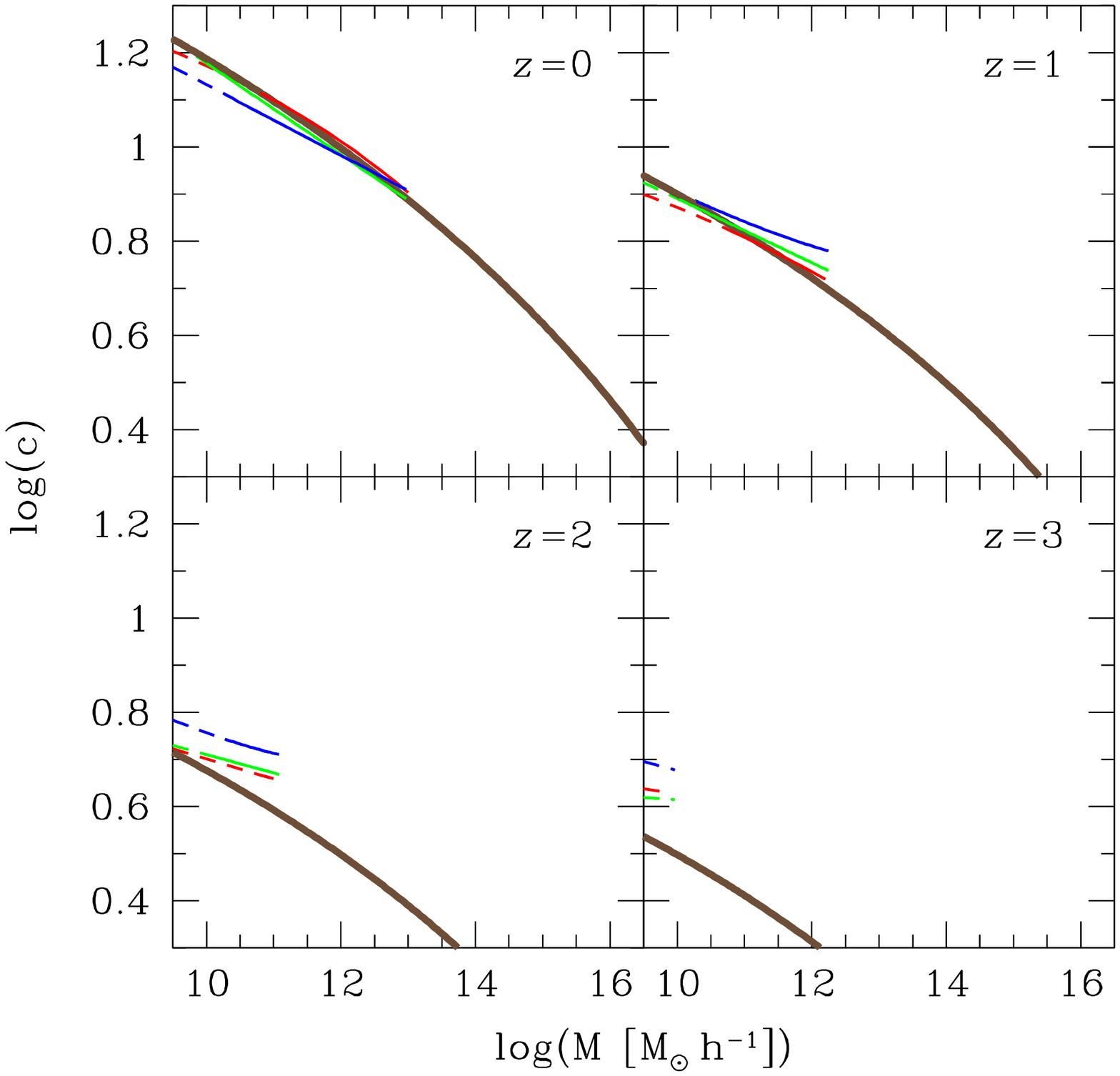}}
 \caption{{\it Top panels}: Comparison between the NFW \Mcb relation predicted by CUSP and provided by the toy models by \citet{Zetal09}, \citet{Mea11} and \citet{Kea11} at different redshifts for $M\vir$ masses in essentially the same {\it WMAP7} cosmology. {\it Bottom panels}: Same as top panels, but for the toy models restricted to masses $M<10 M_\ast(z)$. (A colour version of this Figure is available in the online version of this Journal.)}
 \label{mc1}
\end{figure}

\begin{figure}
\centerline{\includegraphics[scale=0.41,bb=45 150 540 700]{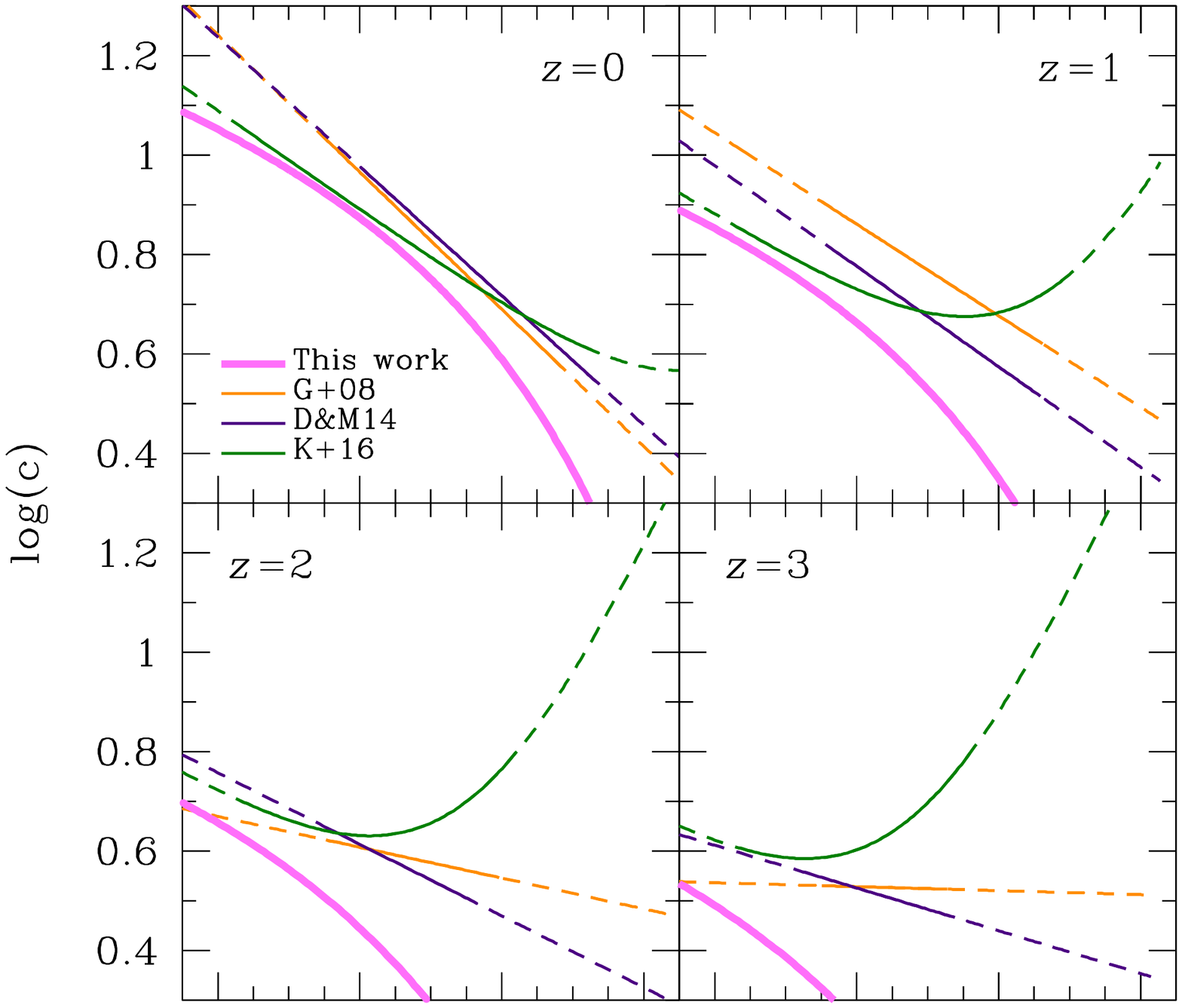}}
\centerline{\includegraphics[scale=0.41,,bb=45 150 540 620]{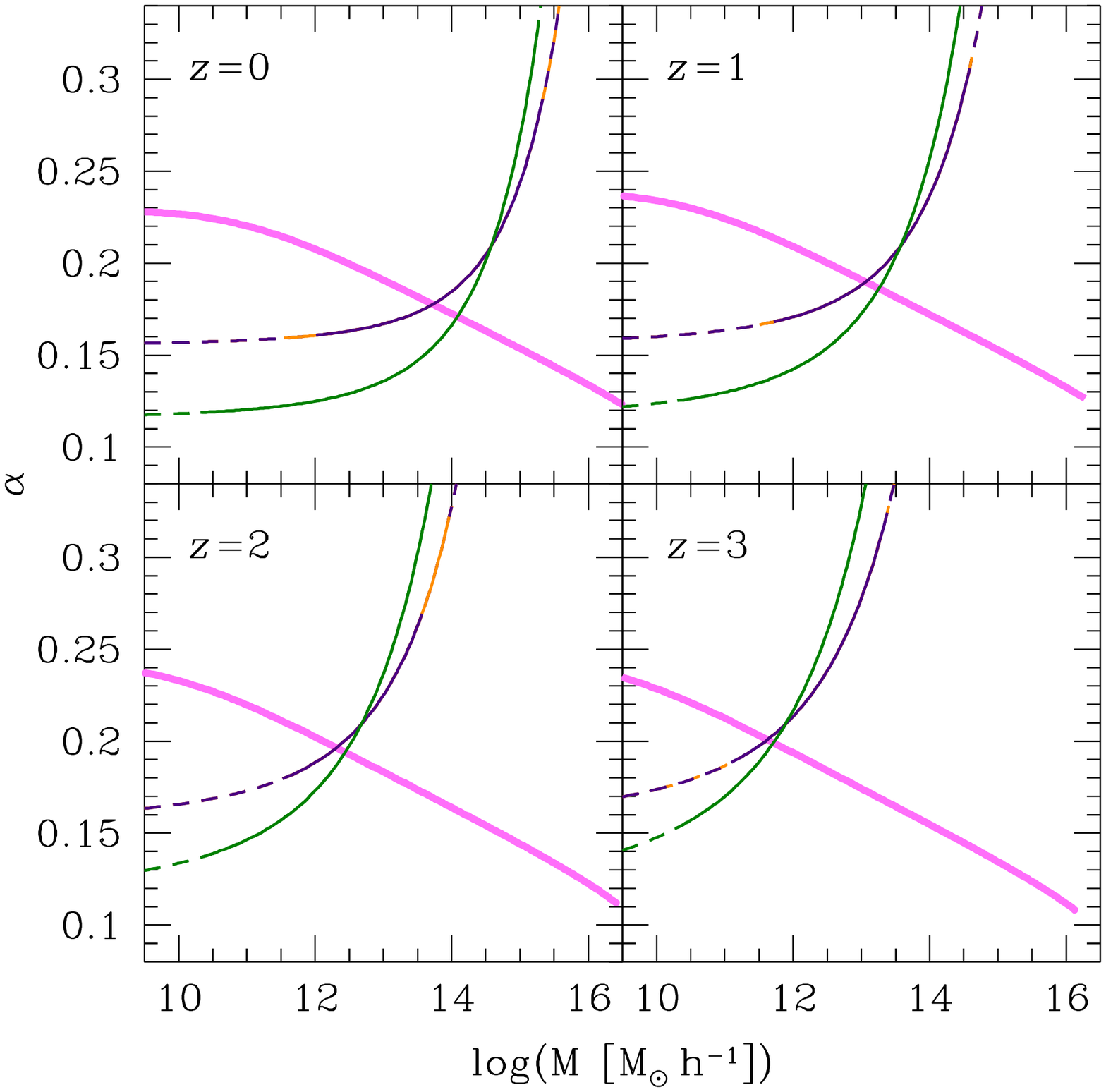}}
 \caption{{\it Top panels}: Same as the top panels of Figure \ref{mc1} for the Einasto \Mcb relation predicted by CUSP and the toy models by \citet{Gea08}, \citet{DM14} and \citet{Kea16} at different redshifts for $M_{\rm 200}$ masses in several {\it Plank14}-like cosmologies. {\it Bottom panels}: Same as top panels for the \Mab relations. (A colour version of this Figure is available in the online version of this Journal.)}
\label{E_mc1}
\end{figure}

As the fit to the three-parametric Einasto function is somewhat degenerate, the simulations employed to build the latter models use many more particles per halo than in the NFW case so as to better determine the halo density profiles. But then halo samples have substantially larger lower mass limits, which makes it difficult to obtain reliable \Mab relations. To alleviate this problem \citet{Gea08} replace this relation between $\alpha$ and $M$ by another one between $\alpha$ and the time-invariant linear (top-hat) height $\nu\F(M,z)\equiv \delta\F\cc(z)/\sigma\F(M,z)$ of protohaloes with $M$ at $z$, which compresses the scatter in the data. Of course, this procedure does not break the degeneracy in $\alpha$; it just smooths out the relation. The price to pay for this is that any real trend in the data is harder to detect. \citet{DM14} adopted the same relation found by \citet{Gea08} and \citet{Kea16} just repeated the fit. Thus, the fact that the \Mab curves derived from the $\nu\F-\alpha$ relations inferred from all three authors essentially coincide does not make them more reliable. On the contrary, they are the relations worst determined. Fortunately, the uncertainty in $\alpha$ has little effect on the associated \Mcb relation (e.g. \citealt{Gea08}; see also Sec.~\ref{physical}).

In all Figures below, the curves predicted by CUSP, from now on called `theoretical relations', are plotted in thick solid line even though there should be essentially no halo in equilibrium with masses $M> 10^3 M_\ast(z)$ where the typical time elapsed since the last major merger is smaller than a few ($2-3$) crossing times; \citealt{Rea01}). Regarding the curves of the toy models, from now on called `empirical relations', they are plotted in thin solid line within the mass range covered by the data, and in thin dashed line their extrapolations beyond that mass range.

\subsubsection{Global Relations}

The theoretical and empirical NFW \Mcb relations are compared in Figure \ref{mc1}, top panel. At $z=0$ all the curves are quite similar. In particular, the theoretical relation fully coincides with the toy model by \citealt{Zetal09}). However, they markedly differ at the high-mass end. While the theoretical curve keeps on decreasing at the same accelerated rate, the empirical curves change their trends unexpectedly: the curve corresponding to the toy model by \citet{Zetal09} suddenly levels off and those of the toy models by \citet{Mea11} and \citet{Kea11} keep on decreasing at a constant rate and bend upwards, respectively. The same divergent behaviour of the toy models is observed at higher redshifts, though at progressively smaller masses. Only when the curves are truncated at $10 M_\ast(z)$ are they much similar to each other (see Fig.~\ref{mc1}, bottom panel) and to our predictions. 

In the Einasto case (see Fig.~\ref{E_mc1}) the result is similar except for the fact that there is a more marked discrepancy between the theoretical predictions and the toy models as well as between the toy models themselves. This is likely due to the fact that, for the above mentioned reasons, the empirical curves are restricted to more massive haloes than in the NFW case, which are the most affected by the departure from equilibrium. Only the toy model by \citet{Kea16} reaches moderately massive haloes and the corresponding \Mcb curves are indeed the closest to the theoretical ones. But the largest discrepancy is in the \Mab relations. While $\alpha$ is little dependent on $M$ and $z$ in the theoretical curves, it markedly depends on both arguments in the empirical ones. We recall that there is a large degeneracy in the $\alpha$ parameter so that the \Mab relation is quite uncertain. The fact that the empirical \Mab curves are very similar to each other does not mean they are more reliable than the theoretical ones. They are similar simply because \citet{DM14} adopted the same \Mab relation as \citet{Gea08} and \citet{Kea16} fitted their own data with identical $\nu\F-\alpha$ functionality. Interestingly, at $z=0$ the theoretical \Mab curve is consistent, between $10^{11}$ \modotb h$^{-1}$ and $10^{15}$ \modotb h$^{-1}$, with a constant value of $\alpha$ of about $0.18$ as found by \citet{LEtal16} in their simulations restricted to that mass range and redshift. 

To sum up, for haloes with masses $M< 10 M_\ast(z)$ at $z\la 2$ there is good agreement between the theoretical global relations and previous toy models targeting masses $M>10^{10} h^{-1}$ \modot. These results indicate that $10 M_\ast(z)$ mark the upper mass limit at each redfhift for halo samples not to be affected by the departure from equilibrium of those objects. The comparison regarding the \Mab relation in the Einasto case is little compelling due to the big uncertainty affecting this relation in the toy models considered and the marked departure from equilibrium of very massive haloes as included in the toy models dealing with the Einasto profile. \citet{Kea11} found, indeed, that the fit of the density profiles of haloes out of equilibrium leads to higher values of $\alpha$ than in relaxed haloes, where it is close to $\alpha \approx 0.18$ (see also \citealt{Kea16}), which agrees with the results by \citet{LEtal16} at $z=0$ and with our predictions.

\begin{figure}
\centerline{\includegraphics[scale=0.41,bb=45 150 540 700]{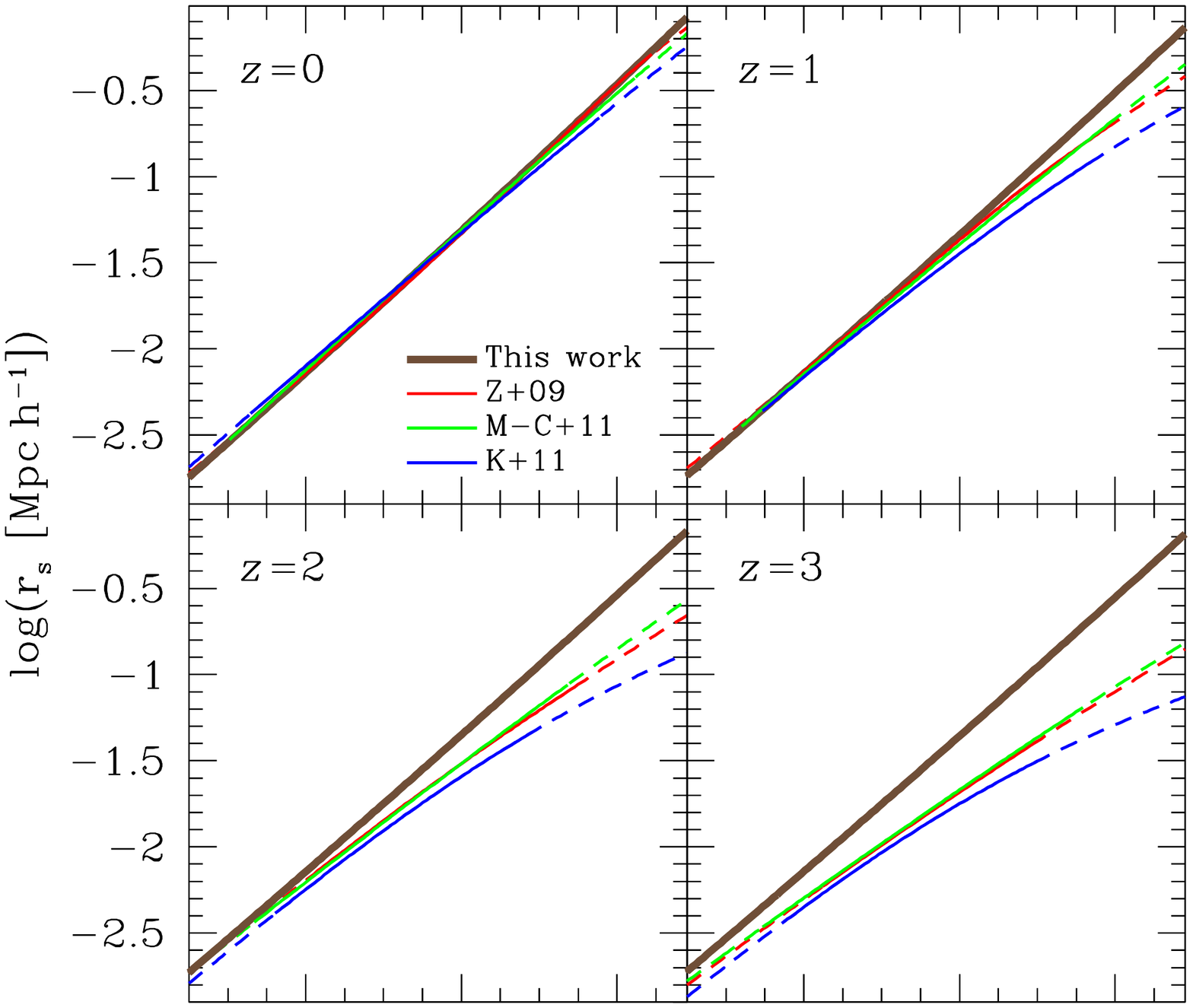}}
\centerline{\includegraphics[scale=0.41,bb=45 150 540 620]{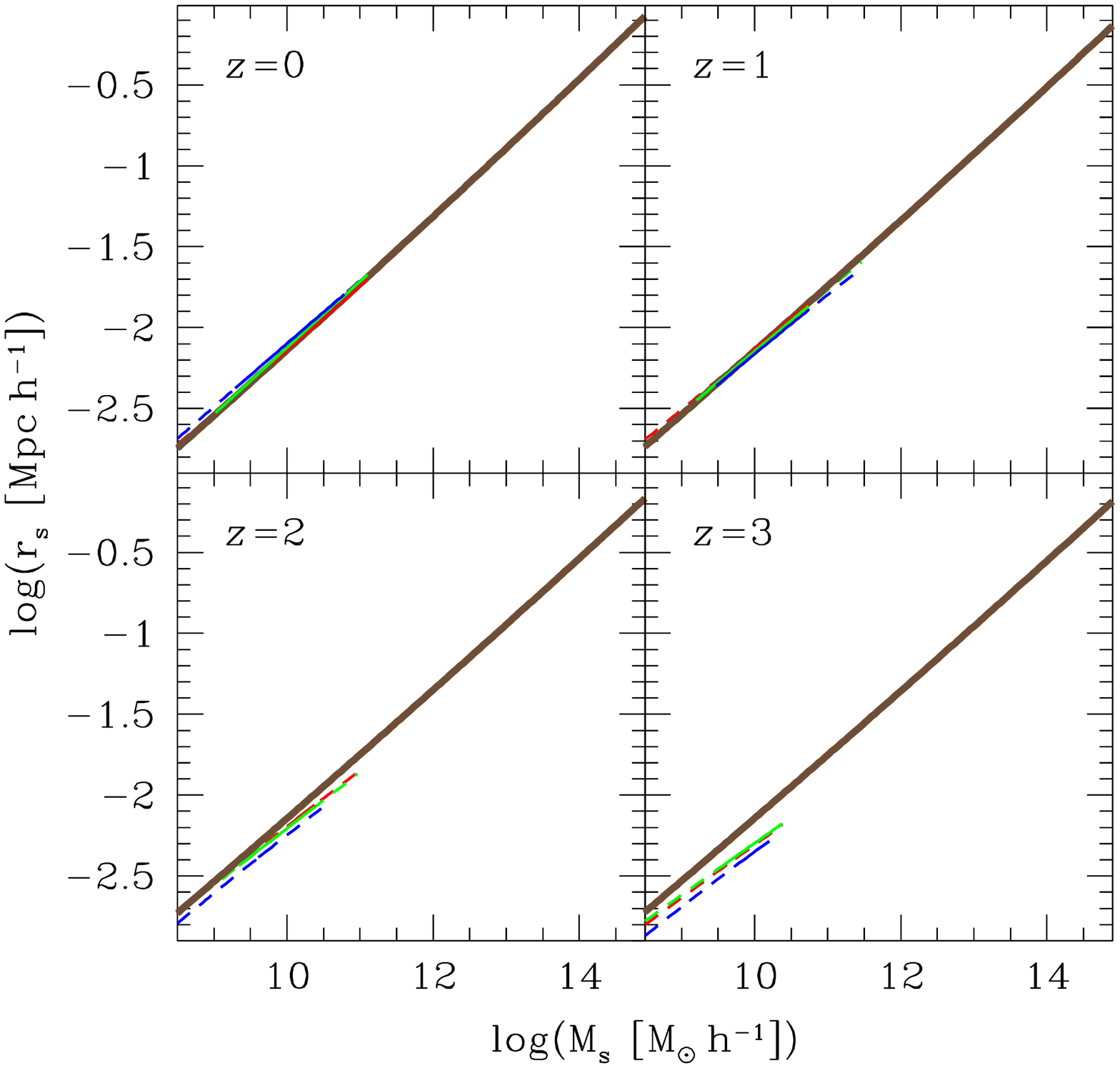}}
\caption{{\it Top panels}: Same as Figure \ref{mc1}, but for the NFW \Msrsb relations. {\it Bottom panels}: Same as top panels, but restricted to haloes with masses $M< 10 M_\ast(z)$. (A colour version of this Figure is available in the online version of this Journal.)}
\label{msrs}
\end{figure}

\begin{figure}
\centerline{\includegraphics[scale=0.41,bb=45 150 540 700]{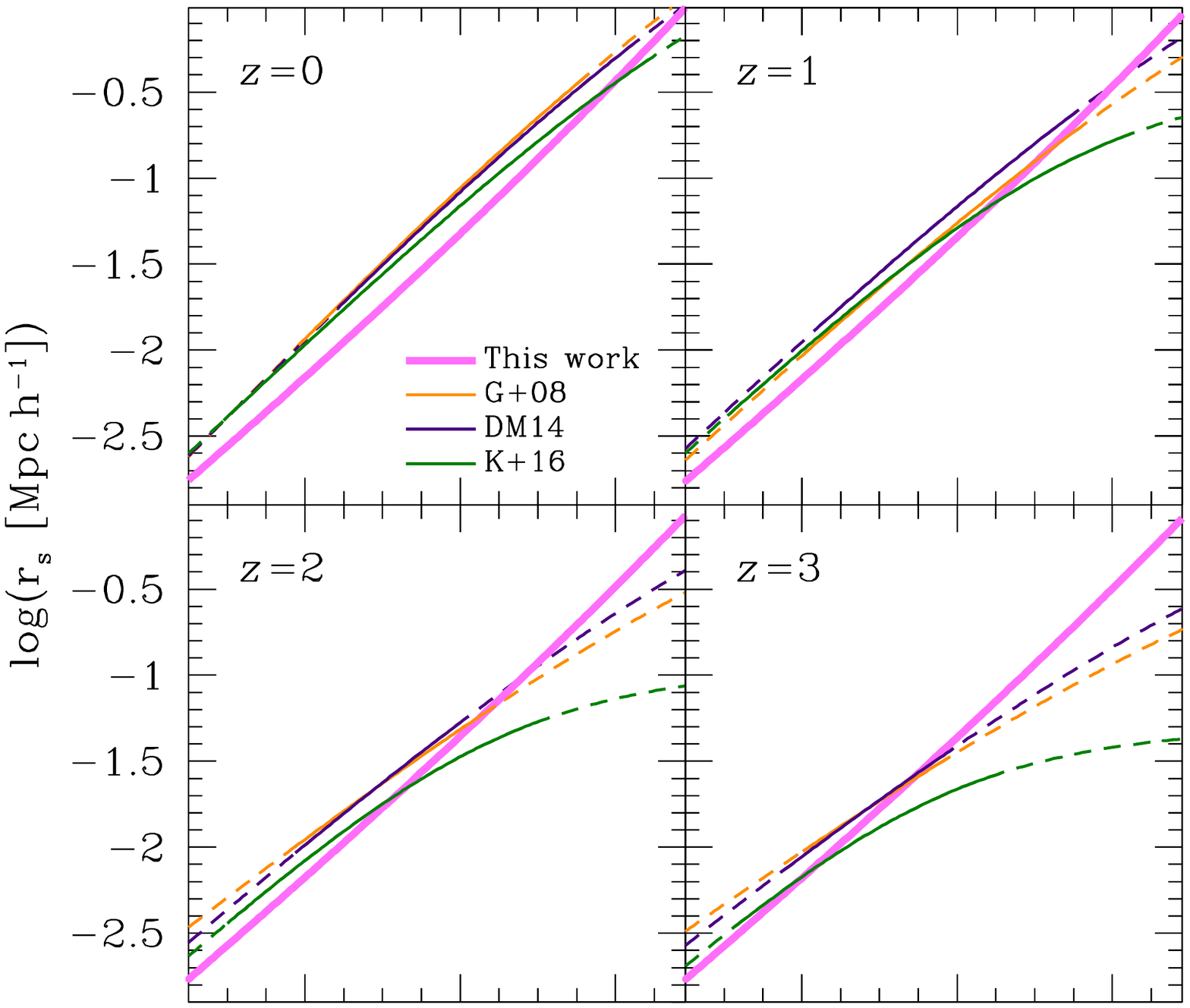}}
\centerline{\includegraphics[scale=0.41,bb=45 150 540 620]{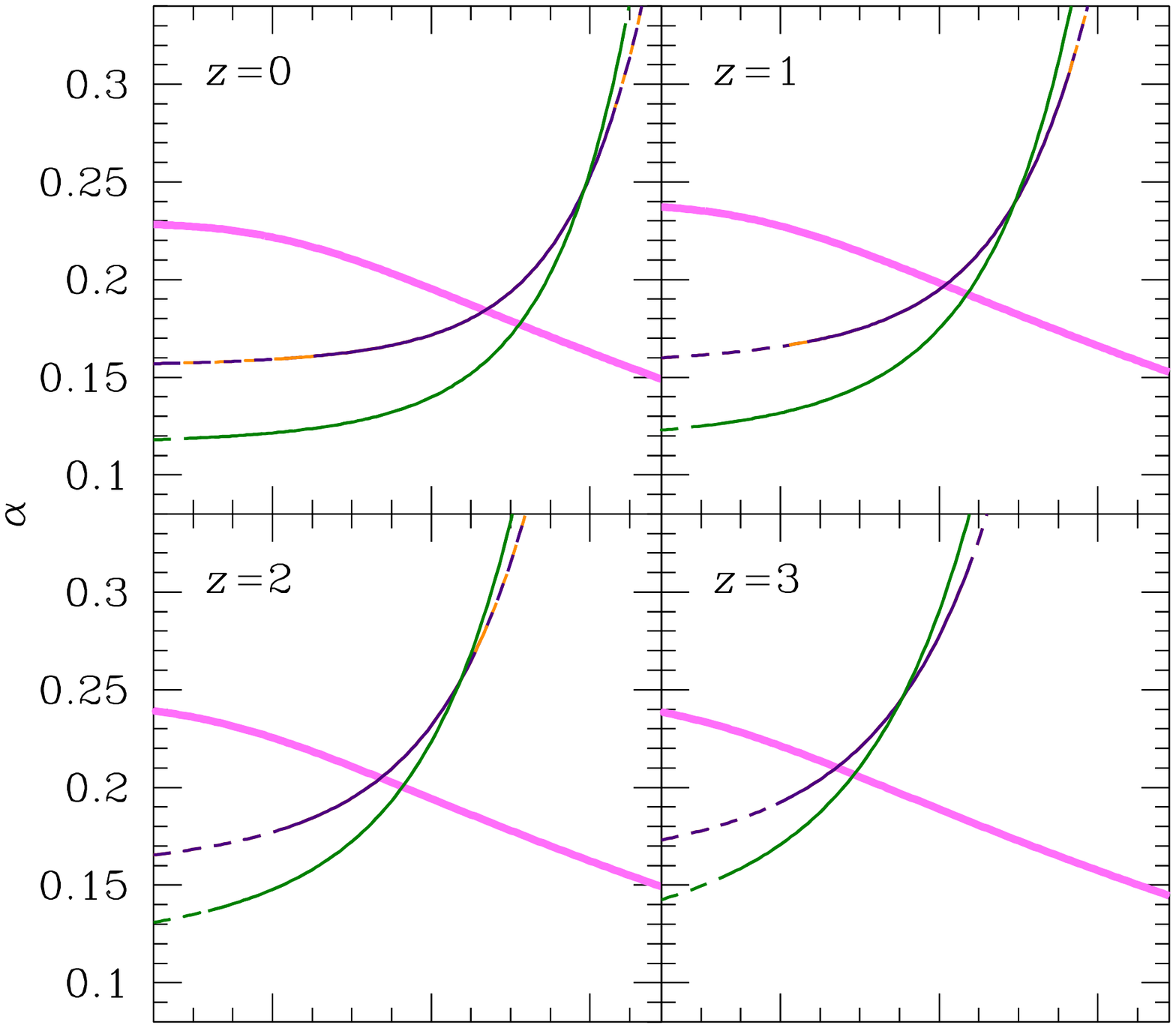}}
\centerline{\includegraphics[scale=0.41,bb=45 150 540 620]{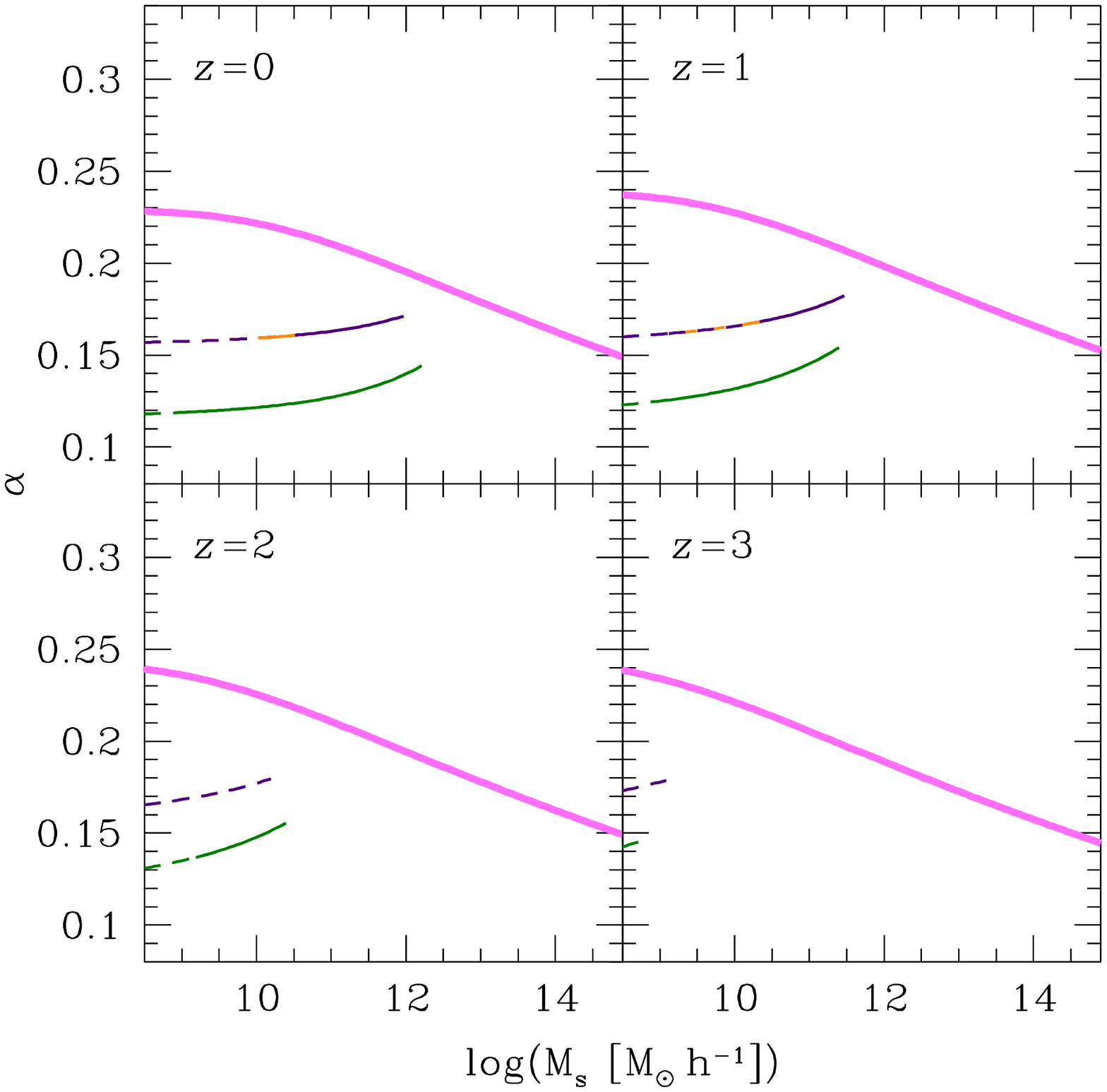}}
\caption{{\it Top panels}: Same as Figure \ref{E_mc1}, but for the \Msrsb relations. {\it Middle panels}: Corresponding \Msab relations. {\it Bottom panels}: Same as middle panels, but restricted to haloes with $M < 10 M_\ast(z)$. (A colour version of this Figure is available in the online version of this Journal.)}
\label{E_msrs}
\end{figure}

\subsubsection{Internal Relations}\label{second_test}

In Figure \ref{msrs}, top panel, we see that, at $z=0$, all NFW \Msrsb curves essentially overlap in the same approximate power-law relation. As $z$ increases, the logarithmic slope of the empirical curves slightly shifts at the high-mass end where it varies notably from author to author. This suggests that the different behaviour of the theoretical an empirical curves at high-masses at every $z$ simply reflects the above mentioned bias introduced by haloes out of equilibrium. Indeed, when the curves are truncated at $10 M_\ast(z)$ (Fig.~\ref{msrs}, bottom panel), not only do they all essentially coincide with each other but also with the thoeretical curves. On the contrary, the slope of the theoretical relation remains much more constant with varying redshift, as expected. In fact, the only slight change disappears when the fit is carried over the same radial range at all redshifts. 

The Einasto relations are shown in Figure \ref{E_msrs}. The theoretical \Msrsb and \Msab curves for different $z$ now almost coincide even without taking a fixed fitting radial range (see top and middle panels). This reflects the fact that the Einasto function provides better fits to the halo density profiles than the NFW function due to the extra parameter $\alpha$ (see Fig.~\ref{r-2}). On the contrary, the empirical curves show a marked dependence on $z$, even more marked than for the NFW profile (Fig.~\ref{msrs}), likely due to the slight coupling of $\rs$ with the poorly determined $\alpha$ parameter, whose wrong dependence on $z$ artificially boosts that of $\rs$. Once again, when the relations are truncated at $10 M_\ast(z)$, all the \Msrsb curves almost fully overlap and their dependence on $z$ disappears (we have skipped this figure to avoid being repetitive). However, the corresponding empirical \Msab curves, depicted in Figure \ref{E_msrs} bottom panel, still get apart from the theoretical one and show a marked dependence on $z$. 

The conclusion of the comparison of the internal relations is that, in the mass range $M\la 10 M_\ast(z)$ where most haloes are in equilibrium, the toy models behave as predicted by CUSP: the \Msrsb relation is close to a time-invariant power-law, and the dependence of $\alpha$ on $\Ms$ is much less marked and closer to constant than found at large masses, according to the predictions of CUSP. Since it is very unlikely that these results are simultaneously met for other causes, they give strong support to the halo growth conditions evidenced by CUSP. Unfortunately, the mass range (of two orders of magnitude) and the redshift interval (below $z=2$ only) covered by those toy models are too narrow to be more conclusive.

\subsection{Whole Mass Range at Redshift Zero}\label{physical}

This limitation is amply overcome by the simulation recently performed by WBFetal at $z=0$ in a flat Lambda 100 Gev WIMP universe with the {\it Plank14} cosmological parameters (see Table \ref{T2}). These authors fitted the empirical \Mcb relation found for $M_{200}$ masses to the Einasto relation (with unconstrained fits). We can thus check the validity of the Einasto CUSP-based analytic expressions derived for those cosmology and mass definition over the whole mass range. 

WBFetal studied two cases: with and without free-streaming mass cut-off of the CDM power spectrum. For simplicity, we concentrate here on the case of no cut-off, though CUSP can also deal with a mass cut-off (see \citealt{Vea12}). To this end we have carried out unconstrained fits to the density profiles predicted by CUSP in the same cosmology and for the same mass definition and derived the analytic \Mcb and \Mab relations (see the resulting values of the coefficients in Tables \ref{T3} and \ref{T4}). 

In Figure 12 we depict the $M_{200}$--$\alpha$ relation obtained from those fits. Like in Figure \ref{E_ma2}, the best values of $\alpha$ show large oscillations, though its trend is quite well determined anyway, particularly in the range $-3 \la \log(M_{200}/$M$_\odot)\la 7$. Nevertheless, the values of the other two parameters, $\Ms$ (or $\rhos$) and $\rs$, are very well determined. This is important because, even though our $\alpha$ values deviate from those around $0.16$ found by WBFetal,\footnote{The reason for that difference could be due to the fact that WBFetal fit the staked density profiles in a different range of radii above an unspecified (possibly mass-dependent) `convergence radius' larger than $10^{-3}R_{200}$.} the corresponding \Msrsb relation is very stable and insensitive to the exact $\alpha$ values. Indeed, as shown in Figure 13, the raw $\rs$ and $\Ms$ values arising from the unconstrained fits with oscillating $\alpha$ values are essentially identical to those found by fixing $\alpha$ according to the analytic expression (\ref{alpha2}) or taking it fixed and equal to 0.22. Indeed, a constant $\alpha$ as adopted by WBFetal is indeed also a good approximation since the absolute variation in the value of this parameter over 15 orders of magnitude (between the maximum at $\log(M_{200})\sim 9$ and the low-mass end at $\log(M_{200})\sim -6$) is only of $\sim 0.02$. As can be seen, the two theoretical \Msrsb relations obtained in that way are almost identical and close to a straight line over more than 20 orders of magnitude, as expected. 

\begin{figure}
\begin{center}
\includegraphics[scale=0.45,bb= 40 150 540 700]{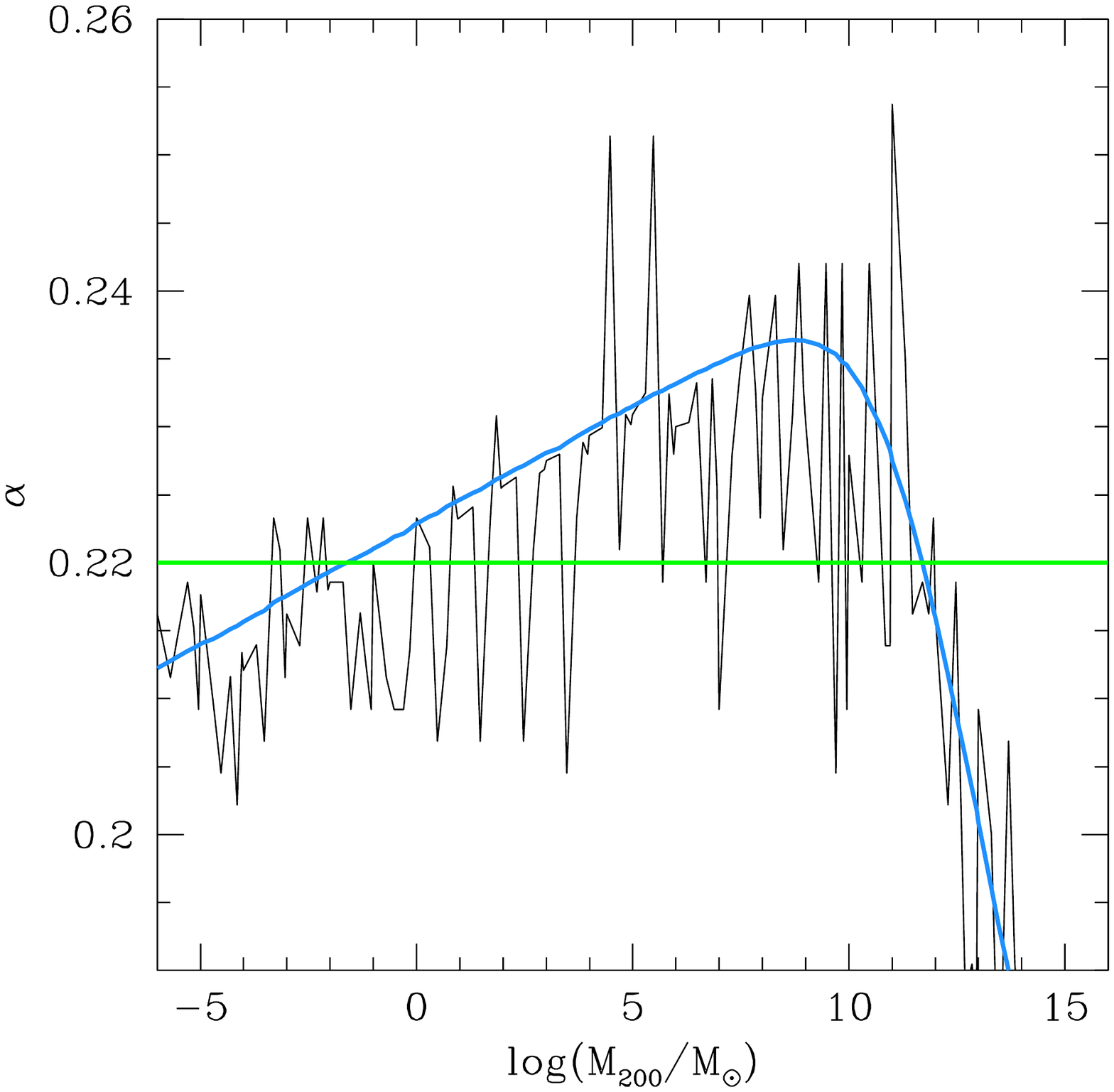}
 \caption{Raw $M_{200}$--$\alpha$ relation (black line) resulting from unconstrained fits to the Einasto function of the density profiles of current haloes predicted by CUSP (with no free-streaming mass cut-off) over the whole mass range analysed by WBFetal in the same cosmology. We also plot the best $\alpha(M_{200})$ fit according to the analytic expression (\ref{alpha2}) with $\Ms=Mf(1)/f(c)$ (blue line) and a constant $\alpha$ value of $0.22$ (green line). To avoid overlapping with the raw relation, the blue curve has been shifted upwards by 0.001. (A colour version of this Figure is available in the online version of this Journal.)}
 \end{center}
\label{new3}
\end{figure}

\begin{figure}
\begin{center}
\includegraphics[scale=0.45,bb= 40 150 540 700]{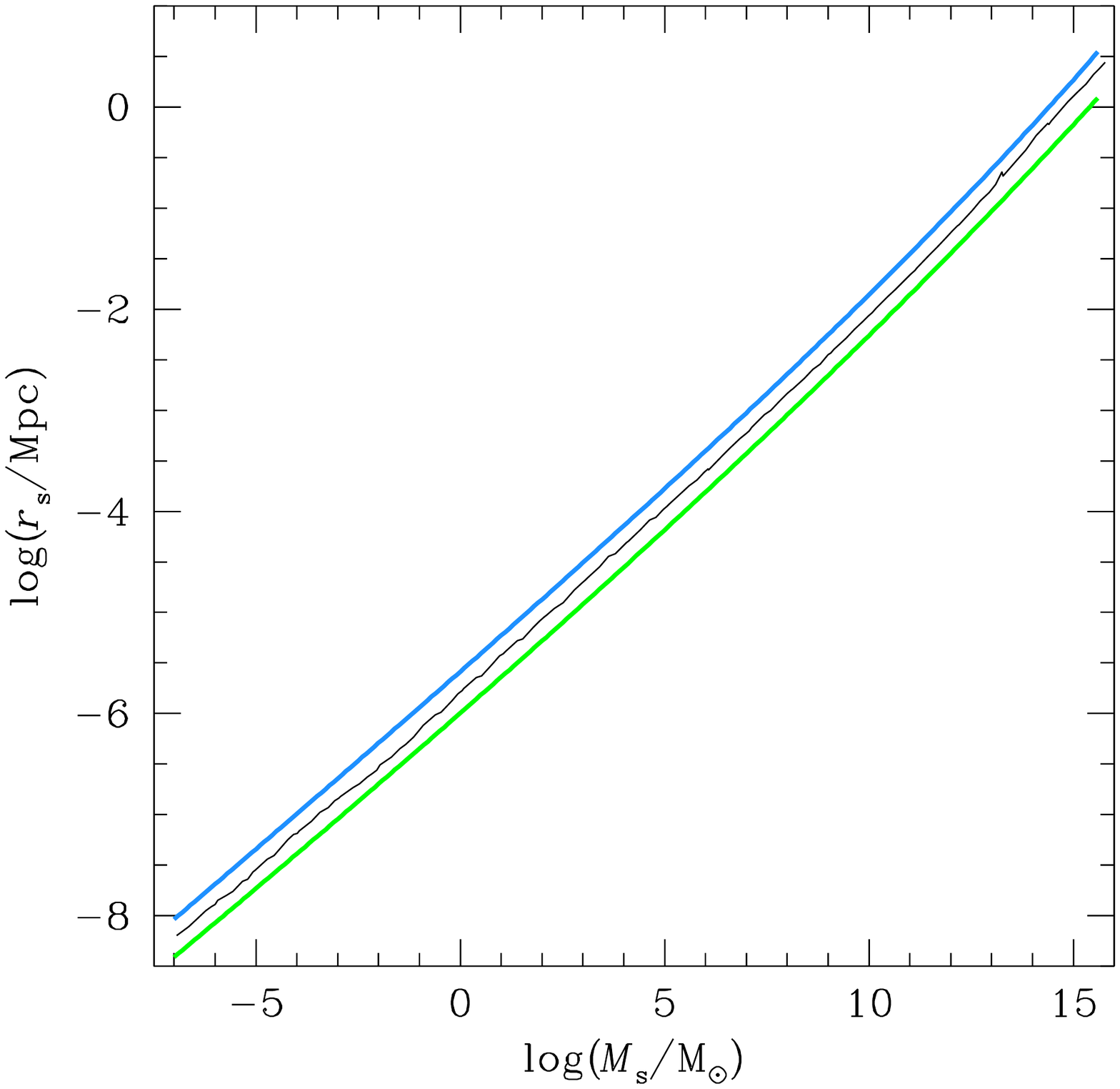}
 \caption{Same as Figure 12, but for the $\rs$ and $\Ms$ values resulting from the unconstrained (triparameteric) fits to the density profiles predicted by CUSP (black line) and from the constrained (biparametric) fits using the two $\alpha(M_{200})$ relations plotted in Figure 12: the one according to the expression (\ref{alpha2}) (blue line) and the other one with $\alpha=0.22$ (green line). To avoid overlapping these two latter curves have been shifted 0.2 dex upwards and downwards, respectively. (A colour version of this Figure is available in the online of this Journal.)}
 \end{center}
\label{new4}
\end{figure}

The corresponding \Mcb relations are depicted in Figure 14 where they are compared to the empirical \Mcb relation found by WBFetal from the fits to the density profiles of simulated haloes (with fixed $\alpha$ equal to 0.16). As can be seen, there is good agreement between both \Mcb relations: the largest difference between the two curves over more than 20 orders of magnitude at $M_{200}\sim 10^{9}$M$_\odot$ is just a factor $\sim 1.15$ and much smaller than the rms scatter of the empirical $c$ values. That agreement is particularly remarkable given that the WBFetal \Mcb relation was obtained by linking by hand the relations obtained in a mosaic of 8 narrow mass ranges which do not exactly match each other and even substantially deviate from the general trend adopted (see their Fig.~3), meaning that this empirical relation could locally somewhat deviate from the real relation.

In Figure 14 we also depict the \Mcb relations obtained from several phenomenological or toy models. Among all those \Mcb relations, the ones showing a global trend similar to that of the WBFetal relation are the phenomenological models by \citet{Cea15}, \citet{LEtal14} (in its latest version given in \citealt{LEtal16}) and, at a lesser extent, by \citet{DJ19}. As mentioned, all these models rely on (or are consistent with) the fact that haloes grow inside-out during accretion as found in CUSP. But what about their treatment of major mergers? Do they also implicity assume that the density profile arising from major mergers is indistinguishable from that of haloes grown by smooth accretion?

\begin{figure*}
\begin{center}
\includegraphics[scale=0.75,bb= 40 250 540 565]{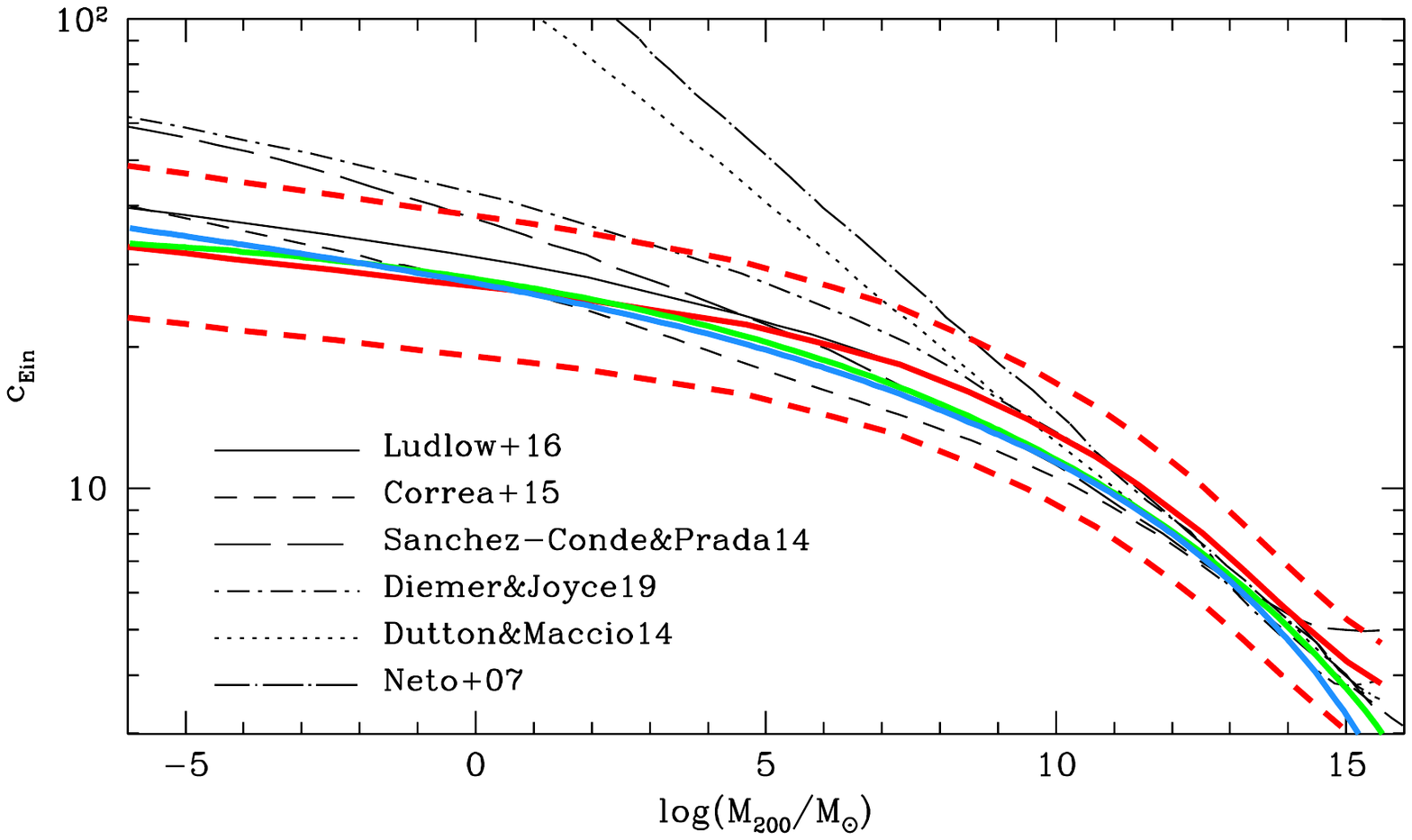}
 \caption{Same as Figure 13 (same lines and colours), but for the $M_{200}$--$c$ relation the blue and green lines are not shifted now and almost overlap). For comparison we plot the $M_{200}$--$c$ relation (for median $c$ values) found by WBFetal by fitting the density profiles of simulated haloes to the Einasto profile with a fixed value of $\alpha$ of 0.16 (solid red line) and the corresponding typical rms scatters (dashed red lines). The \Mcb relations predicted by other phenomenological and toy models are also shown (black lines). (A colour version of this Figure is available in the online version of this Journal.)}
 \end{center}
\label{new6}
\end{figure*}

The equality between the mass profile and MAH of haloes at the base of \citet{LEtal14} model rigorously holds for purely accreting haloes only. When a halo undergoes a major merger its mass suddenly increases by a factor of about two, while the mean (or critical) cosmic density does not change. Major mergers thus cause discontinuities in the halo MAHs that are not reflected in their mass profiles which are necessarily continuous. There should thus be a small trend for halo MAHs to decline slightly more steeply than their mass profiles at high-$z$ where major mergers are more frequent (e.g. \citealt{Zetal03}). That trend is indeed observed in the comparison between the two kinds of profiles made in \citet{LEtal14} (see their Fig.~4). Fortunately, this effect is expected to only affect the density profile derived from the MAH at very small radii, typically smaller than $\rs$, so it should not affect the \Mcb relation derived in this way. Only for haloes with very low masses near the free-streaming cut-off where essentially the whole density profile is set during the initial rapid growth phase (with very frequent major mergers) should this effect have noticeable consequences for the \Mcb relation derived from MAHs. But \citealt{LEtal16} changed their model in that mass regime. Instead of monitoring MAHs, they monitor the history of the collapsed mass, i.e. all the mass that is eventually assembled in the final object regardless of whether it comes from the main progenitor. When doing this, these authors implicitly follow what would be the putative MAH of the final halo had it been evolving by {\it pure accretion} thanks to the fact that, as considered in CUSP, the density (and mass) profile of haloes emerging from major mergers is indistinguishable from that of purely accreting haloes. It is thus unsurprising that the model by \citealt{LEtal16} shows a similar good behaviour than the CUSP-based model. The \citealt{LEtal16} is just slightly less accurate because it uses the EPS formalism instead of CUSP. Moreover, it includes two free parameters, while the CUSP-based model is parameter-free. 

In the models by \citet{Cea15} and \citet{DJ19}, the inside-out growth implied by the constancy of the scale radius is only seen at the late phase of their evolution. In \citet{Cea15} model this is not important because the model does not rely on whether accreting haloes grow inside-out or not, it just relies on the mass aggregation history predicted by the EPS model without making the distinction between smooth accretion and major mergers. However, the \citet{DJ19} model explicitly assumes that, during the late accretion (or pseudo-accretion) phase haloes grow inside-out by keeping the $\rs$ unchanged. In the initial phase dominated by major mergers, it is assumed that it is the concentration $c$ which is kept approximately constant. This approximation not only complicates the model (it harbours 5 free parameters because of the necessity to define the frontier between the two different growth phases dependent on halo mass), but it is not accurate enough. If $c$ were really kept constant in major mergers, $\Ms$ would be kept essentially proportional to $M_{200}$ (eq.~[\ref{new}]) and, since $R_{200}$ is proportional to $M_{200}$ to the 1/3, $\rs$ should also be proportional to $\Ms$ to the 1/3. It is true that, according to the predictions of CUSP, $\tau$ is not far from 1/3 (see the value of $\tau_0$ in Tab.~3), which explains that the \citet{DJ19} model yields acceptable predictions. However, it slightly deviates from this value depending on mass (and redshift), which causes the \Mcb relation predicted by the \citet{DJ19} model not to flatten enough. 

The conclusion of the comparison over the full halo mass range is thus that the predictions of CUSP agree with the results of numerical simulations over the full mass range of 20 orders of magnitude at $z=0$. The fact that the phenomenological models of second generation that recover the flattening of the \Mcb relation include implicitly the inside-out growth of accreting haloes and implicitly the similarity of the density profiles of haloes regardless of their assembly history gives strong support to such growth conditions explicitly accounted for in CUSP (and proven in \citealt{SM19}).

\section{Summary and Conclusions}\label{summ}

The CUSP formalism allows one to accurately derive from first principles and with no free parameter all macroscopic halo properties (including substructure; \citealt{I,II,III}) and to clarify the origin of their features \citep{SM19}. In this Paper it has been applied to derive the mass-scale relations satisfied by halo density profiles. Specifically, we have analysed how the two fundamental characteristics of halo growth evidenced by CUSP, namely that accreting haloes grow inside-out and that haloes having suffered major mergers are indistinguishable from those having grown by pure accretion, translate into those relations. 

We have shown that such characteristics lead to an intrinsic \Mtrtb relation of the real non-parametric density profiles that is time-invariant and very close to a power-law with index $\tau$ around 1/3. However, the proxy relation \Msrsb and the global shape parameter $\alpha$ found from the fit of the profiles to the usual NFW and Einasto parametric functions slightly deviate from those simple trends due to the fact that these functionalities do not yield a perfect fit, so, even though accreting haloes grow inside-out, the best fitting values of the internal and shape parameters, $\rs$, $\Ms$ and $\alpha$, slightly shift with mass and redshift as the total fitted radial range expands. 

We have showed that, while the Einasto function gives acceptable fits to the halo density profile over the whole mass and redshift range, the NFW function is only acceptable, at low-$z$, for high-masses. Simple analytic expressions have been provided that give very good fits to the ``internal'' \Msrsb and \Msab relations as well as to the ``global'' \Mcb and \Mab ones for haloes of all masses and redshifts obtained from the fitting to the NFW and Einasto functions of the non-parametric halo density profiles predicted by CUSP. Even though the two kinds of relations are equivalent, the former are more practical given their simpler form. In particular, the \Mcb relation is far from a power-law as found in some classical toy models since it progressively flattens in log-log towards low-masses. On the other hand, it shows a marked dependence on redshift which differs from a power-law of $1+z$ as also found in some phenomenological models.  

The performance of our CUSP-based analytic \Mcb and \Msrsb relations and the associated \Mab and \Msab ones in the Einasto case has been compared to that of several toy models holding at high masses ($M\ga 10^{10} h^{-1}$ \modot) and low redshifts ($z\la 2$) as well as to several phenomenological models supposed to cover all halo masses. We find good agreement between the predicted \Mcb and \Msrsb relations and the toy models provided halo masses stay below $10 M_\ast(z)$ at any redshift. At higher masses the agreement deteriorates due to the fact that the predictions of CUSP are for virialised haloes, whereas simulated haloes with higher masses progressively get apart from equilibrium (an increasing fraction of them have suffered a too recent major merger and have had no time to relax). Regarding the \Mab and \Msab relations, our predictions substantially deviate from those found by \citet{Gea08} at very high masses where haloes are out of equilibrium. They are, however, consistent with a roughly constant value of $\alpha\sim 0.18$ as found by \citet{LEtal16}.

On the other hand, we have found good agreement with the empirical Einasto \Mcb relation recently derived by WBFetal from a simulation of haloes at $z=0$ with masses spanning more than 20 orders of magnitude. The relations predicted by CUSP behave slightly better than any other phenomenological model put forward so far including those of \citet{Cea15}, \citet{LEtal16} and, to a lesser extent, \citet{DJ19}. We have shown that the latter models, which also behave reasonably well, also implicitly assume the above mentioned fundamental characteristics of halo growth accounted for by CUSP and proven in \citet{SM19}. These characteristics were also assumed in the old phenomenological model by \citet{metal03} using of the EPS formalism. However, the new mass-scale relations derived here from the CUSP formalism are more accurate and practical and arise from first principles, i.e. they do not rely on any arguable assumption and do not use any free parameter.

\vspace{0.75cm} \par\noindent
{\bf ACKNOWLEDGEMENTS} \par

\noindent This work was funded by grants CEX2019-000918-M (Unidad de Excelencia `Mar\'ia de Maeztu') and PID2019-109361GB-100 (together with FEDER funds) by MCIN/AEI/10.13039/501100011033 and by the grant 2017SGR643 by the Catalan DEC.

\par\vspace{0.75cm}\noindent
{\bf DATA AVAILABILITY}

\vspace{11pt}\noindent 
The data underlying this article will be shared on reasonable request to the corresponding author.


\appendix


\section{Median concentration and mean-profile concentration}\label{App1}

The mean density profile of haloes of a given virial or $M_{200}$ mass $M$ and the corresponding radius $R$ is, like the density profile of individual haloes of that mass, approximately of the NFW or Einasto form. Thus, according to equations (\ref{ms1}) or (\ref{ms2}) and equation (\ref{new}), the characteristic density of the mean density profile, $\rhos(\lav\rho\rav)$, equal to the mean density of individual haloes at the scale radius $\rs(\lav\rho\rav)$, satisfies the relation 
\begin{equation}
M=f[R/\rs(\lav\rho\rav)] C \rhos(\lav\rho\rav) \rs^3(\lav\rho\rav),
\end{equation}
where $C$ is a constant equal to $16\pi$ and $2 \pi (2/\alpha)^{1-{\frac{3}{\alpha}}}\,\exp(2/\alpha)$ in the cases of the NFW and Einasto profiles, respectively, and $f(x)$ is the corresponding function. On the other hand, the characteristic density $\rhos$ at the scale radius $\rs$ of each individual halo satisfies the same relation
\begin{equation}
M=f[R/\rs] C \rhos \rs^3.
\end{equation}
We thus have
\begin{equation}
\Delta \ln \rs = \frac{1}{3} \left(\ln \left\{\frac{f[R/\rs(\lav\rho\rav)]}{(R/\rs)}\right\}-\Delta \ln \rhos\right),
\end{equation}
with $\Delta \ln \rs=\ln \rs - \ln[\rs(\lav\rho\rav)]$ and $\Delta \ln \rhos=\ln \rhos - \ln \rhos(\lav\rho\rav)$. Taking into account the relation 
\begin{equation}
\ln \rhos=\ln \rhos(\lav\rho\rav)+\frac{\der \ln\rho}{\der\ln r}\bigg|_{\ln \rs} \Delta \ln\rs,    
\end{equation}
valid to first order, where, by definition of scale radius, the logarithmic derivative in the right-hand member is equal to $-2$, we arrive at
\begin{equation}
\Delta \ln \rs= \ln \left\{\frac{f[R/\rs(\lav\rho\rav)]}{f(R/\rs)}\right\}+\frac{\Delta \rhos}{\rhos(\lav\rho\rav)}, 
\label{fin}
\end{equation}
where $\Delta \rhos=\rho[\rs(\lav\rho\rav)]-\rhos(\lav\rho\rav)$. Note that $\rho[\rs(\lav\rho\rav)]$ is the density of each individual halo at the scale radius of the mean density profile.

Taking into account that $f(x)$ is a very smooth function of $x$, the term $\ln \{f[R/\rs(\lav\rho\rav)]/f(R/\rs)\}$ in equation (\ref{fin}) can be neglected. Thus, dividing $\rs$ and $\rs(\lav\rho\rav)$ by $R$, equation (\ref{fin}) can be rewritten in the form
\begin{equation}
\frac{c}{c(\lav\rho\rav)}\approx \exp{\left[\frac{\Delta \rhos}{\rhos(\lav\rho\rav)}\right]}. 
\label{lognorm}
\end{equation}
Since $c/c(\lav\rho\rav)$ is lognormally distributed \citep{DM14}, equation (\ref{lognorm}) implies that $\Delta \rhos/\rhos(\lav\rho\rav)$ is (approximately) normally distributed. Moreover, since the mean of the latter variable is null, we conclude that the median of $c/c(\lav\rho\rav)$ is $\exp(0)=1$ or, equivalently, that the median concentration of haloes with $M$ very nearly coincides with the concentration of the mean density profile, $c(\lav\rho\rav)$. 


\begin{thebibliography}{}
%
\bibitem[\protect\citeauthoryear{Anderhalden \& Diemand}{2013}]{AD13} Anderhalden D., Diemand J., 2013, JCAP, 2013, 009
%
\bibitem[\protect\citeauthoryear{Ascasibar, Hoffman, \& Gottl{\"o}ber}{2007}]{Aea07} Ascasibar Y., Hoffman Y., Gottl{\"o}ber S., 2007, MNRAS, 376, 393
%
\bibitem[\protect\citeauthoryear{Avila-Reese et al.}{1999}]{AR99} Avila-Reese V., Firmani C., Klypin A., Kravtsov A.~V., 1999, MNRAS, 310, 527
%
\bibitem[Bardeen et al.(1986)]{BBKS} Bardeen J.~M., Bond J.~R., Kaiser N., Szalay A.~S., 1986, \apj, 304, 15
%
\bibitem[\protect\citeauthoryear{Bhattacharya et al.}{2013}]{Bea13} Bhattacharya S., Habib S., Heitmann K., Vikhlinin A., 2013, ApJ, 766, 32
%
\bibitem[Bond et al.(1991)]{BCEK} Bond, J.R., Cole, S., Efstathiou, G., \& Kaiser, N.\ 1991, \apj, 379, 440 (BCEK)
%
\bibitem[Bower(1991)]{B91} Bower R.~G., \mnras, 248, 332
%
\bibitem[Bryan \& Norman(1998)]{bn98} Bryan G.L. \& Norman M.~L., 1998, \apj, 495, 80
%
\bibitem[Bullock et al.(2001)]{Bea01} Bullock J.~S., Kolatt, T.~S., Siga Y. et al., 2001, \mnras, 321, 559 
%
\bibitem[\protect\citeauthoryear{Chen et al.}{2020}]{Cea20} Chen Y., Mo H.~J., Li C., Wang H., Yang X., Zhang Y., Wang K., 2020, ApJ, 899, 81
%
\bibitem[\protect\citeauthoryear{Child et al.}{2018}]{Cea18} Child H.~L., Habib S., Heitmann K., Frontiere N., Finkel H., Pope A., Morozov V., 2018, ApJ, 859, 55
%
\bibitem[\protect\citeauthoryear{Col{\'\i}n et al.}{2004}]{Cea04} Col{\'\i}n P., Klypin A., Valenzuela O., Gottl{\"o}ber S., 2004, ApJ, 612, 50
%
\bibitem[Correa et al.(2015)]{Cea15} Correa C.~A., Wyithe J.~S.~B., Schaye J., Duffy A.~R., 2015, \mnras, 452, 1217
%
\bibitem[\protect\citeauthoryear{Cuesta et al.}{2008}]{Cea08} Cuesta A.~J., Prada F., Klypin A., Moles M., 2008, MNRAS, 389, 385
%
%
\bibitem[\protect\citeauthoryear{Diemand, Kuhlen, \& Madau}{2007}]{DKM07} Diemand J., Kuhlen M., Madau P., 2007, ApJ, 667, 859
%
\bibitem[\protect\citeauthoryear{Diemer, Kravtsov \& More}{2013a}]{Dea13} Diemer B., Kravtsov A.~V., More S., 2013, ApJ, 779, 159
%
\bibitem[\protect\citeauthoryear{Diemer, More \& Kravtsov}{2013b}]{Dea13b} Diemer B., More S., Kravtsov A.~V., 2013, ApJ, 766, 25
%
\bibitem[\protect\citeauthoryear{Diemer \& Kravtsov}{2015}]{DK15} Diemer B., Kravtsov A.~V., 2015, ApJ, 799, 108
%
\bibitem[\protect\citeauthoryear{Diemer \& Joyce}{2019}]{DJ19} Diemer B., Joyce M., 2019, ApJ, 871, 168
%
\bibitem[\protect\citeauthoryear{Dolag et al.}{2004}]{Dea04} Dolag K., Bartelmann M., Perrotta F., Baccigalupi C., 
Moscardini L., Meneghetti M., Tormen G., 2004, \aap, 416, 853
%
\bibitem[Duffy et al.(2008)]{Duf08} Duffy A.~R., Schaye J., Kay S.~T., Dalla Vecchia C., 2008, \mnras, 390, L64
%
%
\bibitem[Dutton \& Macci\`o(2014)]{DM14} Dutton A.~A. \& Macci\`o A.~V., 2014, \mnras, 441, 3359
%
\bibitem[Eke et al.(2001)]{Eea01} Eke V.~R., Navarro J.~F., Steinmetz M., 2001, \apj, 554, 114
%
\bibitem[Einasto(1965)]{E65} Einasto J. 1965, Trudy Inst. Astrofiz. Alma-Ata, 5, 87
%
\bibitem[Fakhouri \& Ma(2009)]{FM09} Fakhouri O. \& Ma, C.-P., 2009, \mnras, 394, 1825
%
\bibitem[Fakhouri \& Ma(2010)]{FM10} Fakhouri O., Ma C.-P., 2010, \mnras, 401, 2245
%
\bibitem[\protect\citeauthoryear{Fukushige \& Makino}{2001}]{FM01} Fukushige T., Makino J., 2001, ApJ, 557, 533
%
\bibitem[\protect\citeauthoryear{Gao et al.}{2008}]{Gea08} Gao, L., Navarro, J.~F., Cole, S., et al.\ 2008, \mnras, 387, 536
%
%
\bibitem[Gottl{\"o}ber, Klypin \& Kravtsov(2001)]{Gea01} Gottl{\"o}ber S., Klypin A., Kravtsov A.~V., 2001, \apj, 546, 223 
%
\bibitem[Gottl{\"o}ber et al.(2002)]{Gea02} Gottl{\"o}ber S., Kerscher  M., Kravtsov A.~V., et al., 2002, \aap, 387, 778 
%
\bibitem[Hahn et al.(2009)]{Hea09} Hahn O., Porciani C., Dekel A., Carollo C.~M., 2009, \mnras, 398, 1742 
%
\bibitem[\protect\citeauthoryear{Hester \& Tasitsiomi}{2010}]{HT10} Hester J.~A., Tasitsiomi A., 2010, ApJ, 715, 342
%
\bibitem[\protect\citeauthoryear{Heitmann et al.}{2015}]{Hea15} Heitmann K., Frontiere N., Sewell C., Habib S., Pope A., Finkel H., Rizzi S., et al., 2015, ApJS, 219, 34
%
%
\bibitem[\protect\citeauthoryear{Hellwing et al.}{2021}]{Hea20} Hellwing W.~A., Cautun M., van de Weygaert R., Jones B.~T., 2021, PhRvD, 103, 063517
%
\bibitem[\protect\citeauthoryear{Henry}{2000}]{H00} Henry, J.~P., 2000, \apj, 534, 565
%
\bibitem[\protect\citeauthoryear{Huss, Jain, \& Steinmetz}{1999}]{Hea99} Huss A., Jain B., Steinmetz M., 1999, ApJ, 517, 64
%
\bibitem[\protect\citeauthoryear{Ishiyama et al.}{2013}]{Iea13} Ishiyama T., Rieder S., Makino J., Portegies Zwart S., Groen D., Nitadori K., de Laat C., et al., 2013, ApJ, 767, 146
%
\bibitem[Ishiyama(2014)]{I14} Ishiyama, T.\ 2014, \apj, 788, 27
%
\bibitem[\protect\citeauthoryear{Ishiyama et al.}{2020}]{Iea20} Ishiyama T., Prada F., Klypin A.~A., Sinha M., Metcalf R.~B., Jullo E., Altieri B., et al., 2020, arXiv, arXiv:2007.14720
%
\bibitem[\protect\citeauthoryear{Juan et al.}{2014a}]{Jea14a} Juan E., Salvador-Sol\'e E., Dom\`enec G., Manrique A., 2014a, \mnras, 439, 719
%
\bibitem[Juan et al.(2014b)]{Jea14b} Juan E., Salvador-Sol{\'e} E., Dom{\`e}nech G., Manrique A., 2014b, \mnras, 439, 3156
%
\bibitem[\protect\citeauthoryear{Klypin, Trujillo-Gomez \& Primack}{2011}]{Kea11} Klypin A.~A., Trujillo-Gomez S., Primack
  J., 2011, \apj, 740, 102
%
\bibitem[Klypin et al.(2016)]{Kea16} Klypin, A., Yepes, G., Gottl{\"o}ber, S., Prada, F., He{\ss}, S.\ 2016, \mnras,  
%
\bibitem[\protect\citeauthoryear{Komatsu et al.}{2011}]{Km11} Komatsu E., Smith K. M., Dunkley J., Bennet C. L., Gold B., Hinshaw G.,
  Jarosik N., Larson D., and 13 others, 2011, \apjs, 192, 18 
%
\bibitem[Lacey \& Cole(1994)]{LC94} Lacey, C., \& Cole, S.\ 1994, \mnras, 271, 671
%
\bibitem[\protect\citeauthoryear{Loeb \& Peebles}{2003}]{LP03} 
Loeb A., Peebles P.~J.~E., 2003, ApJ, 589, 29
%
\bibitem[\protect\citeauthoryear{Lu et al.}{2006}]{Luea06} Lu Y., Mo H.~J., Katz N., Weinberg M.~D., 2006, \mnras 368, 1931 
%
%
\bibitem[\protect\citeauthoryear{Lud\l ow et al.}{2013}]{LEtal13} 
  Lud\l ow A.~D., Navarro J.~F., Boylan-Kolchin M., et al.\ 2013, \mnras, 432, 1103 
%
\bibitem[\protect\citeauthoryear{Lud\l ow et al.}{2014}]{LEtal14} 
  Lud\l ow A.~D., Navarro J.~F., Angulo R.~E., et al., 2014, \mnras, 441, 378 
%
\bibitem[\protect\citeauthoryear{Lud\l ow et al.}{2016}]{LEtal16} Lud\l ow A.~D., Bose S., Angulo R.~E., Wang L., Hellwing W.~A., Navarro J.~F., Cole S., et al., 2016, MNRAS, 460, 1214
%
\bibitem[\protect\citeauthoryear{Macci\`o et al.}{2008}]{Mea08}
  Macci\`o A.~V., Dutton A.~A., van den Bosch F.~C., 2008, \mnras 391, 1940
%
\bibitem[\protect\citeauthoryear{Manrique \& Salvador-Sol\'e}{1995}]{MSS95} 
  Manrique A. \& Salvador-Sol\'e E., 1995, \apj, 453, 6
%
\bibitem[\protect\citeauthoryear{Manrique \& Salvador-Sol\'e}{1996}]{MSS96} 
  Manrique A. \& Salvador-Sol\'e E., 1996, \apj, 467, 504
%
\bibitem[Manrique et al.(1998)]{Mea98} Manrique, A., Raig, A., Solanes, J.~M., et al.\ 1998, \apj, 499, 548 
%
\bibitem[\protect\citeauthoryear{Manrique et al.}{2003}]{metal03} Manrique A., Raig A., Salvador-Sol\'e E., Sanchis T., Solanes J.~M., 2003, \apj, 593, 26
%
\bibitem[\protect\citeauthoryear{Mao, Zentner, \& Wechsler}{2018}]{Mea18} Mao Y.-Y., Zentner A.~R., Wechsler R.~H., 2018, MNRAS, 474, 5143
%
%
\bibitem[\protect\citeauthoryear{Moore et al.}{2001}]{Mea01} Moore B., Calc{\'a}neo-Rold{\'a}n C., Stadel J., Quinn T., Lake G., Ghigna S., Governato F., 2001, PhRvD, 64, 063508
%
\bibitem[\protect\citeauthoryear{Mu{\~n}oz-Cuartas et al.}{2011}]{Mea11} Mu{\~n}oz-Cuartas J.~C., Macci{\`o} A.~V., Gottl{\"o}ber S., Dutton, A.~A., 2011, \mnras, 411, 584
%
\bibitem[\protect\citeauthoryear{Navarro, Frenk \& White}{1995}]{NFW95} Navarro J.~F., Frenk C.~S, White S.~D.~M., 1995, \apj, 275, 720
%
\bibitem[Navarro et al.(1996)]{NFW96} Navarro J.~F., Frenk C.~S., White S.~D.~M., 1996, \apj, 462, 563
%
%
\bibitem[\protect\citeauthoryear{Navarro et al.}{2004}]{Navea04} Navarro J.~F., Hayashi E., Power C., Jenkins A. R., Frenk, C. S.,  White, S. D. M., Springel, V., Stadel, J., Quinn, T. R., 2004, \mnras, 349, 1039
%
%
\bibitem[\protect\citeauthoryear{Neto et al.}{2007}]{Nea07} Neto A.~F.,  Gao L., Bett P., et al., \mnras, 3381, 1450
%
\bibitem[Peebles(1980)]{P80} Peebles, P.~J.~E.\ 1980, Large-Scale Structure of the Universe by Phillip James Edwin Peebles. Princeton University Press, 1980. ISBN: 978-0-691-08240-0
%
\bibitem[Planck Collaboration et al.(2014)]{P14} Planck Collaboration, Ade P.~A.~R., Aghanim N., et al., 2014, \aap, 571, AA16
%
\bibitem[\protect\citeauthoryear{Prada et al.}{2012}]{PEtal12} Prada F., Klypin A.~A., Cuesta A.~J., Betancort-Rijo J.~E., Primack J., 2012, \mnras, 423, 3018
%
\bibitem[\protect\citeauthoryear{Press \& Schechter}{1974}]{PS} Press, W.~H., \& Schechter, P.\ 1974, \apj, 187, 425 
%
\bibitem[\protect\citeauthoryear{Raig, Gonz{\'a}lez-Casado, \& Salvador-Sol{\'e}}{2001}]{Rea01} Raig A., Gonz{\'a}lez-Casado G., Salvador-Sol{\'e} E., 2001, MNRAS, 327, 939
%
\bibitem[\protect\citeauthoryear{Ramakrishnan, Paranjape, \& Sheth}{2020}]{Rea20} Ramakrishnan S., Paranjape A., Sheth R.~K., 2021, MNRAS, 503, 2053
%
\bibitem[\protect\citeauthoryear{Romano-D\'\i az et al.}{2006}]{RD06} Romano-D\'\i az E., Faltenbacher A., Jones D., Heller C., Hoffman Y.,
  Shlosman I., 2006, \apj, 637, L93
%
%
\bibitem[\protect\citeauthoryear{Salvador-Sol\'e, Solanes \& Manrique}{1998}]{ssm} Salvador-Sol\'e E., Solanes J.~M., Manrique A., 1998, \apj, 499, 542
%
\bibitem[\protect\citeauthoryear{Salvador-Sol{\'e}, Manrique \& Solanes}{2005}]{Sea05} Salvador-Sol{\'e} E., Manrique A., Solanes J.~M., 2005, \mnras, 358, 901
%
\bibitem[\protect\citeauthoryear{Salvador-Sol\'e et al.}{2007}]{Sea07} Salvador-Sol\'e E., Manrique A., Gonz\'alez-Casado G., Hansen S.~H., 2007, \apj, 666, 181
%
\bibitem[\protect\citeauthoryear{Salvador-Sol\'e et al.}{2012a}]{Sea12a} Salvador-Sol\'e E., Vi\~nas J., Manrique A., Serra S., 2012, \mnras, 423, 2190
%
\bibitem[Salvador-Sol{\'e} et al.(2012b)]{Sea12b} Salvador-Sol{\'e} E., Serra S., Manrique A., Gonz{\'a}lez-Casado G., 2012b, \mnras, 424, 3129
%
\bibitem[\protect\citeauthoryear{Salvador-Sol{\'e} \& Manrique}{2021}]{SM19} Salvador-Sol{\'e} E., Manrique A., 2021, ApJ, 914, 141
%
\bibitem[\protect\citeauthoryear{Salvador-Sol{\'e}, Manrique, \& Botella}{2022a}]{I} Salvador-Sol{\'e} E., Manrique A., Botella I., 2022, \mnras, 509, 5305
%
\bibitem[\protect\citeauthoryear{Salvador-Sol{\'e}, Manrique, \& Botella}{2022b}]{II} Salvador-Sol{\'e} E., Manrique A., Botella I., 2022, \mnras, 509, 5316
%
\bibitem[Salvador-Sol{\'e} et al.(2022)]{III} Salvador-Sol{\'e} E., Manrique A., Canales D., Botella I., 2022, \mnras, 511, 641
%
\bibitem[S{\'a}nchez-Conde \& Prada(2014)]{SP14} S{\'a}nchez-Conde M.~A. \& Prada F., 2014, \mnras, 442, 2271
%
\bibitem[Sheth \& Tormen(2004)]{ST04} Sheth R.~K. \& Tormen G.\ 2004, \mnras, 350, 1385
%
\bibitem[\protect\citeauthoryear{Springle et al.}{2005}]{Sprea05} Springel V., White S.~D.~M., Jenkins A., et al., 2005, Nature, 435, 629
%
\bibitem[\protect\citeauthoryear{Sugiyama}{1995}]{S95} Sugiyama, N.\ 1995, \apjs, 100, 281
%
\bibitem[\protect\citeauthoryear{van den Bosch}{2002}] {vdB02} van den Bosch F. C., 2002, MNRAS, 331, 98
%
\bibitem[\protect\citeauthoryear{van den Bosch et al.}{2014}]{Vea14} van den Bosch F.~C., Jiang F., Hearin A., Campbell D., Watson D., Padmanabhan N., 2014, MNRAS, 445, 1713
%
\bibitem[Vi{\~n}as, Salvador-Sol{\'e} \& Manrique(2012)]{Vea12} Vi{\~n}as J.,  Salvador-Sol{\'e} E., Manrique A., 2012, \mnras, 424, L6 
%
\bibitem[\protect\citeauthoryear{Wang \& White}{2009}]{WW09} Wang J. \& White S.~D.~M., 2009, \mnras, 396, 709
%
\bibitem[\protect\citeauthoryear{Wang et al.}{2011}]{Wea11} Wang J., Navarro J.~F., Frenk C.~S., White S.~D.~M., Springel V., Jenkins A., Helmi A., et al., 2011, MNRAS, 413, 1373
%
\bibitem[\protect\citeauthoryear{Wang et al.}{2020a}]{Wea20a} Wang K., Mao Y.-Y., Zentner A.~R., Lange J.~U., van den Bosch F.~C., Wechsler R.~H., 2020, MNRAS, 498, 4450
%
\bibitem[\protect\citeauthoryear{Wang et al.}{2020}]{Wea20b} Wang J., Bose S., Frenk C.~S., Gao L., Jenkins A., Springel V., White S.~D.~M., 2020, Nature, 585, 39
%
\bibitem[\protect\citeauthoryear{Zhao et al.}{2003}]{Zetal03} Zhao D. H., Mo H. J., Jing Y. P., B\"orner G., 2003, MNRAS, 339, 12
%
\bibitem[\protect\citeauthoryear{Zhao et al.}{2009}]{Zetal09} Zhao, D.~H., Jing, Y.~P., Mo, H.~J., B\"orner, G.\ 2009, \apj, 707, 354 

\end{thebibliography}
\end{document}